\documentclass{lmcs}
\pdfoutput=1

\usepackage{lastpage}
\lmcsdoi{15}{1}{34}
\lmcsheading{}{\pageref{LastPage}}{}{}%
{Jan.~24,~2018}{Mar.~29,~2019}{}

\usepackage[latin1]{inputenc}

\usepackage{amsmath}
\newcommand{\mathlarger}[1]{#1}
\usepackage{tikz}
\usepackage{amssymb}
\usepackage{xcolor}
\usepackage{url}
\usepackage{stmaryrd}
\usepackage{lineno,hyperref}

\theoremstyle{plain}\newtheorem{lemma}[thm]{Lemma}

\newcommand{\IsabelleHOL}{Isabelle/HOL}
\newcommand{\separ}{\hspace{2mm}}

\newcommand{\chessgeneralfont}[1]{\mathit{#1}}
\newcommand{\chessgeneral}[1]{\chessgeneralfont{{#1}}}
\newcommand{\varA}[1]{\chessgeneralfont{#1}} 

\newcommand{\chessendgamefont}[1]{\mathtt{#1}}
\newcommand{\chessendgame}[2][KRK]{\chessendgamefont{#1.#2}} 
\newcommand{\chessendgameEG}[1]{\chessendgame[\llbracket EG \rrbracket]{#1}}
\newcommand{\varI}[1]{\chessendgamefont{#1}} 

\newcommand{\postype}{pos}
\newcommand{\legalpos}{lgl\_pos}
\newcommand{\legalmove}{lgl\_move}
\newcommand{\checkmate}{mate}
\newcommand{\chebyshev}{\chessendgame{chebyshev\_dist}}
\newcommand{\side}{side}
\newcommand{\white}{White}
\newcommand{\black}{Black}
\newcommand{\piece}{piece}
\newcommand{\king}{King}
\newcommand{\queen}{Queen}
\newcommand{\rook}{Rook}
\newcommand{\bishop}{Bishop}
\newcommand{\knight}{Knight}
\newcommand{\sq}{square}
\newcommand{\onturn}{on\_turn}
\newcommand{\onsquare}{on\_square}
\newcommand{\opp}{opp}
\newcommand{\board}{on\_board}
\newcommand{\files}{F}
\newcommand{\ranks}{R}
\newcommand{\empt}{empty}
\newcommand{\occupies}{occupies}
\newcommand{\kingscope}{king\_scope}
\newcommand{\rookscope}{rook\_scope}
\newcommand{\clearline}{clear\_line}
\newcommand{\squarebetween}{sq\_btw}
\newcommand{\squarebetweenhv}{\squarebetween\_hv}
\newcommand{\attacks}{attacks}
\newcommand{\incheck}{in\_chk}

\newcommand{\winning}{Winning}
\newcommand{\firstlegal}{first\_move}
\newcommand{\minlegal}{min\_move}

\newcommand{\ws}{WS}
\newcommand{\bb}{B}

\newcommand{\KRKPosition}{\chessendgamefont{KRKPosition}}
\newcommand{\WhiteOnTurn}[1]{\chessendgamefont{WhiteTurn}\separ #1}

\newcommand{\WK}[1]{\chessendgamefont{WK}\separ #1}
\newcommand{\BK}[1]{\chessendgamefont{BK}\separ #1}
\newcommand{\WRopt}[1]{\chessendgamefont{WRopt}\separ #1}
\newcommand{\WR}[1]{\chessendgamefont{WR}\separ #1}
\newcommand{\WRcaptured}[1]{\chessendgamefont{WRcapt}\separ #1}

\newcommand{\chessendgameI}[1]{\chessendgame{#1}}

\newcommand{\room}{\chessendgameI{room}}

\newcommand{\kingssquare}{\chessendgameI{kings\_move}}
\newcommand{\rookssquare}{\chessendgameI{rooks\_move}}

\newcommand{\WRattacksBK}{\chessendgameI{WR\_attacks\_BK}}
\newcommand{\BKCannotMove}{\chessendgameI{BK\_cannot\_move}}
\newcommand{\WRDivides}{\chessendgameI{WR\_divides}}
\newcommand{\LPattern}{\chessendgameI{L\_pattern}}
\newcommand{\DivideAttempt}{\chessendgameI{divide\_attempt}}

\newcommand{\WRExposed}{\chessendgameI{WR\_exposed}}
\newcommand{\WKcanmoveto}{\chessendgameI{WK\_can\_move\_to}}
\newcommand{\WRcanmoveto}{\chessendgameI{WR\_can\_move\_to}}
\newcommand{\moveWK}{\chessendgameI{move\_WK}}
\newcommand{\moveWR}{\chessendgameI{move\_WR}}
\newcommand{\movewhite}{\chessendgameI{move\_white}}
\newcommand{\StrategyWhiteMove}{\chessendgameI{st\_wht\_move}}
\newcommand{\StrategyWhiteMoveFun}{\chessendgameI{st\_wht\_move\_fun}}
\newcommand{\legalmoveWR}{\chessendgameI{\legalmove\_WR}}
\newcommand{\legalmoveBK}{\chessendgameI{\legalmove\_BK}}
\newcommand{\legalmoveWhite}{\chessendgameI{\legalmove\_white}}
\newcommand{\firstlegalWK}{\chessendgameI{\firstlegal\_WK}}
\newcommand{\firstlegalWR}{\chessendgameI{\firstlegal\_WR}}
\newcommand{\firstlegalWhite}{\chessendgameI{\firstlegal\_white}}
\newcommand{\minlegalWR}{\chessendgameI{\minlegal\_WR}}

\newcommand{\retrograde}{retrograde}

\newcommand{\Pre}{\chessendgamefont{Pre}}
\newcommand{\Seq}{\chessendgamefont{Seq}}
\newcommand{\Post}{\chessendgamefont{Post}}
\newcommand{\M}{\chessendgamefont{M}}

\newcommand{\BasicMoves}{\chessendgameI{BasicMoves}}
\newcommand{\MateMoves}{\chessendgameI{MateMoves}}

\newcommand{\ImmediateMateCond}{\chessendgameI{immediate\_mate\_cond}}
\newcommand{\ReadyToMateCond}{\chessendgameI{ready\_to\_mate\_cond}}
\newcommand{\SqueezeCond}{\chessendgameI{squeeze\_cond}}
\newcommand{\RookSafeCond}{\chessendgameI{rook\_safe\_cond}}
\newcommand{\RookHomeCond}{\chessendgameI{rook\_home\_cond}}

\newcommand{\noImmediateMateWK}{\chessendgameI{no\_mate\_WK}}
\newcommand{\noImmediateMateWR}{\chessendgameI{\chessendgame{no\_mate\_WR}}}
\newcommand{\noImmediateMate}{\chessendgameI{no\_immediate\_mate}}

\newcommand{\noSqueezeMove}{\chessendgameI{no\_squeeze}}
\newcommand{\noReadyToMate}{\chessendgameI{no\_ready\_to\_mate}}

\newcommand{\abstr}[1]{\widehat{#1}}

\newcommand{\rel}[1]{{{#1}_{rel}}}
\newcommand{\ndfun}[1]{{{#1}_{set}}}
\newcommand{\dfun}[1]{{{#1}_{fun}}}

\newcommand{\wsA}{\chessgeneral{\ws}}
\newcommand{\bA}{\chessgeneral{\bb}}
\newcommand{\wsI}{\chessendgame{\ws}}
\newcommand{\bI}{\chessendgame{\bb}}

\newcommand{\bkf}{{\textit{bk}_x}}
\newcommand{\bkr}{{\textit{bk}_y}}
\newcommand{\wkf}{{\textit{wk}_x}}
\newcommand{\wkr}{{\textit{wk}_y}}
\newcommand{\wrf}{{\textit{wr}_x}}
\newcommand{\wrr}{{\textit{wr}_y}}

\newcommand{\pA}{\chessgeneral{\postype}}
\newcommand{\sideA}{\chessgeneral{\side}}
\newcommand{\whiteA}{\chessgeneral{\white}}
\newcommand{\blackA}{\chessgeneral{\black}}
\newcommand{\pieceA}{\chessgeneral{\piece}}
\newcommand{\kingA}{\chessgeneral{\king}}
\newcommand{\queenA}{\chessgeneral{\queen}}
\newcommand{\rookA}{\chessgeneral{\rook}}
\newcommand{\bishopA}{\chessgeneral{\bishop}}
\newcommand{\knightA}{\chessgeneral{\knight}}
\newcommand{\squareA}{\chessgeneral{\sq}}
\newcommand{\onturnA}{\chessgeneral{\onturn}}
\newcommand{\onsquareA}{\chessgeneral{\onsquare}}
\newcommand{\legalposA}{\chessgeneral{\legalpos}}
\newcommand{\legalmoveA}{\chessgeneral{\legalmove}}
\newcommand{\checkmateA}{\chessgeneral{\checkmate}}
\newcommand{\oppA}{\chessgeneral{\opp}}
\newcommand{\boardA}{\chessgeneral{\board}}
\newcommand{\filesA}{\chessgeneral{\files}}
\newcommand{\ranksA}{\chessgeneral{\ranks}}
\newcommand{\emptA}{\chessgeneral{\empt}}
\newcommand{\occupiesA}{\chessgeneral{\occupies}}
\newcommand{\kingscopeA}{\chessgeneral{\kingscope}}
\newcommand{\rookscopeA}{\chessgeneral{\rookscope}}
\newcommand{\clearlineA}{\chessgeneral{\clearline}}
\newcommand{\squarebetweenA}{\chessgeneral{\squarebetween}}
\newcommand{\squarebetweenhvA}{\chessgeneral{\squarebetweenhv}}
\newcommand{\attacksA}{\chessgeneral{\attacks}}
\newcommand{\incheckA}{\chessgeneral{\incheck}}

\newcommand{\bool}{bool}

\newcommand{\pI}{\chessendgame{\postype}}
\newcommand{\pEG}{\chessendgameEG{\postype}}
\newcommand{\initialpositions}{inital\_positions}

\newcommand{\legalposI}{\chessendgame{\legalpos}}
\newcommand{\legalmoveI}{\chessendgame{\legalmove}}
\newcommand{\checkmateI}{\chessendgame{\checkmate}}
\newcommand{\incheckI}{\chessendgame{\incheck}}

\newcommand{\movefont}[1]{\mathtt{#1}}
\newcommand{\MoveKind}{\movefont{MoveKind}}
\newcommand{\ImmediateMateMove}{\movefont{ImmediateMate}}
\newcommand{\ReadyToMateMove}{\movefont{ReadyToMate}}
\newcommand{\SqueezeMove}{\movefont{Squeeze}}

\newcommand{\ApproachNonDiagMove}{\movefont{ApproachNonDiag}}
\newcommand{\ApproachDiagMove}{\movefont{ApproachDiag}}
\newcommand{\KeepRoomNonDiagMove}{\movefont{KeepRoomNonDiag}}
\newcommand{\KeepRoomDiagMove}{\movefont{KeepRoomDiag}}

\newcommand{\RookHomeMove}{\movefont{RookHome}}
\newcommand{\RookSafeMove}{\movefont{RookSafe}}

\newcommand{\movefonttxt}[1]{\textit{#1}}
\newcommand{\ImmediateMateMoveTxt}{\movefonttxt{ImmediateMate}}
\newcommand{\ReadyToMateMoveTxt}{\movefonttxt{ReadyToMate}}
\newcommand{\SqueezeMoveTxt}{\movefonttxt{Squeeze}}
\newcommand{\ApproachMoveTxt}{\movefonttxt{Approach}}
\newcommand{\ApproachNonDiagMoveTxt}{\movefonttxt{ApproachNonDiag}}
\newcommand{\ApproachDiagMoveTxt}{\movefonttxt{ApproachDiag}}
\newcommand{\KeepRoomNonDiagMoveTxt}{\movefonttxt{KeepRoomNonDiag}}
\newcommand{\KeepRoomDiagMoveTxt}{\movefonttxt{KeepRoomDiag}}
\newcommand{\KeepRoomMoveTxt}{\movefonttxt{KeepRoom}}
\newcommand{\RookHomeMoveTxt}{\movefonttxt{RookHome}}
\newcommand{\RookSafeMoveTxt}{\movefonttxt{RookSafe}}
\newcommand{\RookSafeSmallBoardsMoveTxt}{\movefonttxt{RookSafeSmallBoards}}

\newcommand{\LIArookscope}{\chessendgame[LIA]{\rookscope}}

\newenvironment{isar_code}[1]
  {\begin{small} #1 }  { \end{small}}

\begin{document}

\title[Computer-Assisted Proving of Combinatorial Conjectures Over Finite Domains]{Computer-Assisted Proving of Combinatorial 
   Conjectures Over Finite Domains: A Case Study of a Chess Conjecture}

\author[Predrag Jani\v{c}i\'c]{Predrag Jani\v{c}i\'c\rsuper{a}}
\address{\lsuper{a}Faculty of Mathematics, University of Belgrade, Serbia}
\email{\{janicic,filip\}@matf.bg.ac.rs}

\author[Filip Mari\'c]{Filip Mari\'c\rsuper{a}}

\address{\lsuper{b}Faculty of Humanities and Social Sciences, University of Rijeka, Croatia}
\email{marko@ffri.hr}
\author[Marko Malikovi\'c]{Marko Malikovi\'c\rsuper{b}}

\begin{abstract}
  There are several approaches for using computers in deriving
  mathematical proofs. For their illustration, we provide an in-depth
  study of using computer support for proving one complex
  combinatorial conjecture -- correctness of a strategy for the chess
  KRK endgame. The final, machine verifiable result presented in this
  paper is that there is a winning strategy for white in the KRK
  endgame generalized to $n \times n$ board (for natural $n$ greater
  than $3$). We demonstrate that different approaches for
  computer-based theorem proving work best together and in synergy and
  that the technology currently available is powerful enough for
  providing significant help to humans deriving some complex proofs.
\end{abstract}




\maketitle

\section{Introduction}
\label{sec:intro}

Over the last several decades, automated and interactive theorem
provers have made huge advances which changed the mathematical
landscape significantly. Theorem provers are already widely used in
many areas of mathematics and computer science, and there are already
proofs of many extremely complex theorems developed within proof
assistants and with many lemmas proved or checked automatically
\cite{Hales05,ARwithFlyspeck,fmzbornik}. We believe there are changes
still to come, changes that would make new common mathematical
practices and proving process will be more widely supported by tools
that automatically and reliably prove some conjectures and even
discover new theorems. Generally, proving mathematical conjectures can
be assisted by computers in several forms:
\begin{itemize}
\item for exhaustive analysis (e.g., for checking of all possible
  cases);
\item for automated proving of relevant statements (e.g., by generic
  automated provers or by solvers for specific theories);
\item for interactive theorem proving (e.g., for proving correctness
  of exhaustive analysis algorithms or for direct proving of relevant
  statements).
\end{itemize}

Each of the above forms of support provides different sorts of
arguments, each has its limitations, and its strengths and weaknesses.
In this paper, we advocate that it is their synergy that provides a
way for proving some complex combinatorial
conjectures. Namely, in every proving process, a human mathematician
experiments, analyses special cases, tries to discover or prove
simpler conjectures, etc. However, all these are typically hidden in
the final product and published mathematics is typically the art of
polished proofs, rarely the art of how to reach them. Computer support
can be crucial in the demanding process of seeking and proving new
mathematical truths.  However, for each computer-supported proving
approach, one has to consider the following key questions:

\begin{itemize}
 \item What have we {\em really} proved?
 \item Is our proof a real mathematical proof, or just a supporting argument?
 \item How reliable is our proof?
 \item What was the level of automation and the level of human effort
   required to make the proof?
\end{itemize}

As a case study, we use a conjecture from one of favourite domains for
many AI approaches -- chess. We consider a conjecture that states {\em
  correctness of a strategy for one chess endgame}.  Endgame
strategies provide concise, understandable, and intuitive instructions
for the player and correctness means that the strategy always leads to
the best possible outcome (under any play by the opponent). We show
that both chess strategies and proofs of their correctness can be
rigorously formalized, i.e.,~expressed in pure mathematical
terms. Although a strategy for an endgame such as KRK (white king and
white rook against black king) is simple, its formalization has a
number of details and it is not easy to prove its correctness. One of
our goals and contributions is modelling the chess rules and chess
endgames so that the correctness proofs can be made as automated,
efficient and reliable as possible.

Through this complex exercise, we will show that a real-world process
of proving conjectures such as correctness of a chess strategy can be
naturally based on a synergy between different computer-supported
approaches and combines experimentation, testing special cases,
checking properties, exploring counterexamples for some conjectures,
and, at the end, proving conjectures within a proof assistant, using
as much automation as possible. Correctness of the strategy for the
KRK endgame is just an example, and the main purpose of this work is
to illustrate a methodology that can be used in proving some complex
combinatorial conjectures in an efficient and a highly reliable
way. In our previous work, we proved the correctness of a very
similar KRK strategy within a constraint solving system URSA
\cite{ursa} and a proof assistant \IsabelleHOL \cite{krk-cade}. In
this paper, we give a unifying perspective of proving conjectures like
that one using three kinds of systems -- a general purpose programming
language, a constraint solver, and a proof assistant, and we further
improve earlier proofs in terms of efficiency, reliability,
understandability and generality.

\paragraph{Overview of the paper} 
In Section \ref{sec:back} we give some background on automated
  theorem proving, SAT and SMT, constraint programming systems and
  URSA, interactive theorem proving and \IsabelleHOL. In Section
  \ref{sec:representation} we discuss formal representation of chess
  and chess endgames in various logic and languages. In Section
  \ref{sec:reasoning} we discuss different methods for reasoning about
  chess endgames and proving their correctness. In Section
  \ref{sec:advanced} we discuss different methods to make our proofs
  faster, and our conjectures higher-level and more general. In
  Section \ref{sec:related} we discuss some related work and in
  Section \ref{sec:conclusions} we draw final conclusions.

\section{Background}
\label{sec:back}

\subsection{Automated and Interactive Theorem Proving}

\paragraph{Automated theorem proving, SAT and SMT}
Modern automated theorem provers based on uniform procedures, such as
the resolution method, and also specific solvers, such as SAT and SMT
solvers, can decide validity of huge formula coming a wide spectrum of
areas including software and hardware verification, model checking,
termination analysis, planning, scheduling, cryptanalysis,
etc. \cite{SATHandbook}. One of the most widely used SMT theories is
linear arithmetic, a decidable fragment of arithmetic (over integers
-- LIA, or reals -- LRA) that uses only addition -- multiplication is
only allowed by a constant number, and $nx$ is just a shorthand for
$x+x+\ldots+x$ where $x$ occurs $n$ times. Linear arithmetic is rather
simple, but expressible enough to be widely used in applications in
computer science \cite{SATHandbook}. There are several decision
procedures for variants of linear arithmetic and they are widely
available through modern SMT solvers \cite{DutertreM06}.

\paragraph{Constraint programming systems and URSA} Constraint
programming systems allow specifying problems and searching for models
that meet given conditions, by using various approaches (e.g.,
constraint logic programming over finite domains, answer set
programming, disjunctive logic programming). Some constraint systems,
such as URSA \cite{ursa}, are based on reduction to propositional
satisfiability problem (SAT). In URSA, the problem is specified in a
language which is imperative and similar to C, but at the same time,
is declarative, as the user does not have to provide a solving
mechanism for the given problem. URSA allows two types of variables:
(unsigned) numerical (with names beginning with \verb|n|, e.g.,
\verb|nX|) and Boolean (with names beginning with \verb|b|, e.g.,
\verb|bX|), with a wide range of C-like operators (arithmetic,
relational, logical, and bitwise). Variables can have concrete
(ground) or symbolic values (in which case, they are represented by
vectors of propositional formulae). There is support for procedures
and there are control-flow structures (in the style of C). Loops must
be with known bounds and there is no \verb|if-else| statement, but
only \verb|ite| expression (corresponding to \verb|?:| in C). An URSA
specification is symbolically executed and the given constraint
corresponds to one propositional formula. It is then transformed into
CNF and passed to one of the underlying SAT solvers. If this formula
is satisfiable, the system can return all its models.

\paragraph{Interactive theorem proving and \IsabelleHOL} 
Interactive theorem provers or {\em proof assistants} are systems used
to check proofs constructed by the user, by verifying each proof step
with respect to the given underlying logic \cite{SeventeenProvers}.
Proofs written within proof assistants are typically much longer than
traditional, pen-and-paper proofs \cite{Barendregt2005} and are
considered to be very reliable \cite{BarendregtB02}. Modern proof
assistants support a high-level of automation and significant parts of
proofs can be constructed automatically. Some proof assistants are
also connected to powerful external automated theorem provers and SMT
solvers, and thanks to that are now capable of proving very complex
combinatorial conjectures.

Isabelle \cite{isabelle} is a generic proof assistant, but its most
developed application is higher order logic
(\IsabelleHOL). Formalizations of mathematical theories are made by
defining new notions (types, constants, functions, etc.), and proving
statements about them (lemmas, theorems, etc.). This is often done
using the declarative proof language Isabelle/Isar
\cite{isar}. \IsabelleHOL{} incorporates several automated provers
(e.g., classical reasoner, simplifier) and it has been connected to
SMT solvers \cite{isabelle-z3}, enabling users to employ SMT solvers
to discharge some goals that arise in interactive theorem proving.

\subsection{Chess Endgame Strategies}

Techniques used by computer programs for chess in midgames
(minimax-style algorithms) are often not appropriate for endgames and
then other techniques have to be used. One such technique is based on
lookup tables (i.e., endgame databases) with pre-calculated optimal
moves for each legal position. However, such tables for endgames with
more chess pieces require a lot of memory and, in addition, they are
completely useless for human players.\footnote{The Lomonosov Endgame
  Tablebases that contain optimal play for all endgames with seven or
  less pieces, generated by Zakharov and Makhnichev
  (\url{http://tb7.chessok.com/}) have around 140 Terabytes. It was
  shown that there is a position with seven pieces such that black can
  be mated in 545 moves but not in less moves, if she/he plays
  optimally}. One alternative to huge lookup tables, usable both to
human and computer players, are {\em endgame strategies}. Endgame
strategies are algorithms that provide concise, understandable, and
intuitive instructions for the player. Endgame strategies do not need
to ensure optimal moves (e.g., shortest path winning moves), but must
ensure correctness -- i.e., if a player $A$ follows the strategy,
he/she should always reach the best possible outcome. The main focus
of our work is formal analysis of combinatorial algorithms, so we will
consider only endgame strategies (and not endgame databases).

One of the simplest chess endgames is the KRK endgame. There are
several winning strategies for white for this endgame. Some of these
were designed by humans, while some are generated semi-automatically
or automatically, using endgame databases, certain sets of human
advices, and approaches such as inductive logic programming, genetic
programming, neural networks, machine learning, etc. However, only a
few of them are really human-understandable. Some strategies for white
were proposed by Zuidema~\cite{Zuidema1974} (a strategy based on
a high-level advice instead of search),
Bratko~\cite{Bratko78,BratkoKM78,BratkoProlog} (an advice-based
strategy consisting of several sorts of strategic moves), Seidel
\cite{Seidel} (a strategy using the ring structure of the chessboard),
Morales \cite{Morales,Morales96} (short strategies produced by
inductive logic programming assisted by a human), etc.
  
\subsubsection{Bratko-Style Strategy for White for the KRK Endgame}
\label{sec:ourbratko}

In the rest of the paper, we will consider one strategy for white
for KRK: it is a variation of Bratko's strategy and 
slightly modified with respect to the version published earlier 
\cite{icga}. 

We assume standard chess notions such as legal moves, mate, stalemate,
etc.~(as defined in the FIDE Handbook \cite{fide}).  {\em Legal KRK
  positions} contain three pieces: the white king (WK), the white rook
(WR), and the black king (BK). On the $8 \times 8$ board, there are
399\,112 such positions, while the strategy is applied only to
175\,168 of them -- those with white on turn.

\paragraph{Auxiliary Notions}
In the following text, we assume that files (columns) and ranks (rows)
of the chessboard are associated with the numbers $0, 1, \ldots, 7$,
and the squares are represented by $(x, y)$ pairs of natural numbers
between $0$ and $7$. For formulating the strategy, we use several
auxiliary notions (standard notions or notions introduced by Bratko):

\noindent
{\bf Manhattan distance:}
For two squares of the chessboard $(x_1, y_1)$ and $(x_2, y_2)$, the
Manhattan distance equals $|x_1-x_2|+|y_1-y_2|$.

\noindent
{\bf Chebyshev distance:}
For two squares of the chessboard $(x_1, y_1)$ and $(x_2, y_2)$, the
Chebyshev distance is the minimal number of moves a king requires to
move between them, i.e., $\max(|x_1-x_2|, |y_1-y_2|).$

\begin{figure}[h]
\noindent
\begin{center}
\includegraphics[width=2.7cm]{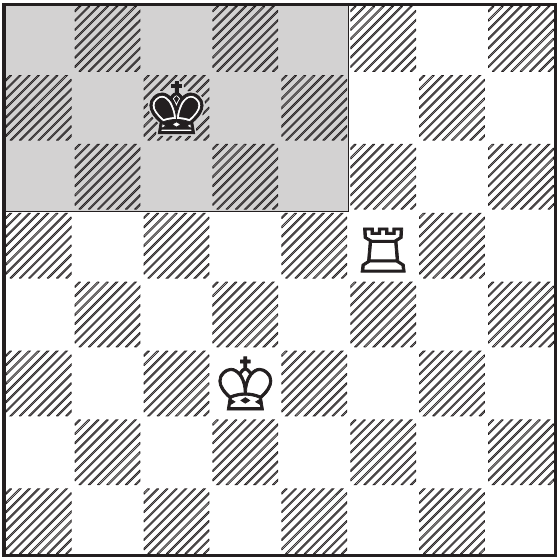}
\hspace*{2mm}
\includegraphics[width=2.7cm]{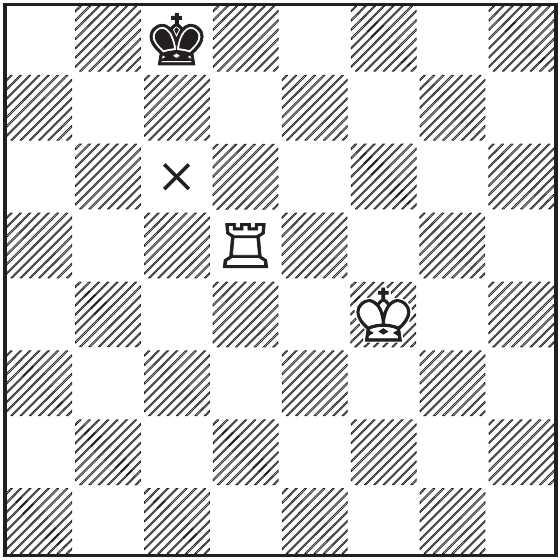}
\hspace*{2mm}
\includegraphics[width=2.7cm]{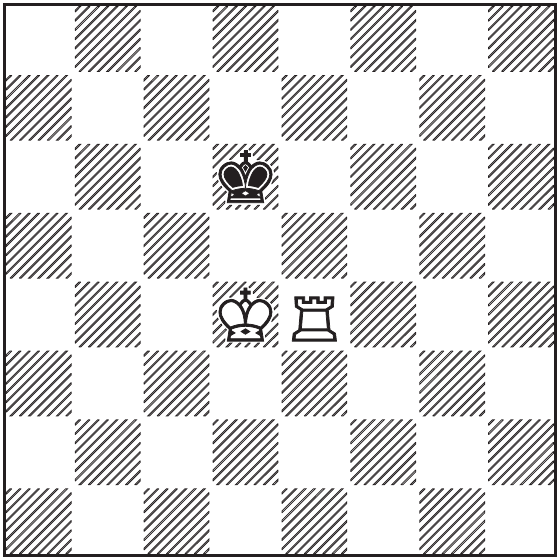}
\hspace*{2mm}
\includegraphics[width=2.7cm]{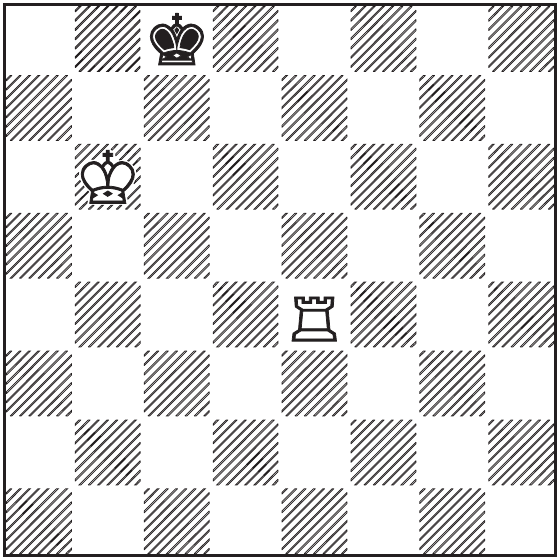}
\end{center}
\caption{From left to right: illustration of the notion of {\em room},
  of the notion of {\em critical square}, of the notion of {\em
    L-pattern}, and of a position in which the $\ReadyToMateMoveTxt$
  move is possible}
\label{fig:room}
\end{figure}

\noindent {\bf Room:} 
Following the strategy, white tries to squeeze the rectangular space
available to black king -- that space is called the {\em room}
(Figure \ref{fig:room}) and is measured by its half-perimeter. When
the black king and the white rook are in line, the black king is not
confined (not restricted to a rectangular area), and the room takes
the value 15 (whenever the black king is confined, the half-perimeter
of the guarded space is at most 14)\footnote{On the $n\times n$ board,
  the half-perimeter is at most $2n - 2$, and when the black king is
  not confined, the room is $2n-1$.}. Therefore, if the rook is on the
square $(\wrf, \wrr)$ and the black king on the square $(\bkf,\bkr)$,
then room equals:

$$\left\{ \begin{array}{ll}
15,          &  \mbox{if $\wrf=\bkf$ or $\wrr=\bkr$}  \\
x + y,       &  \mbox{otherwise}
\end{array} \right.$$

\noindent where $x$ and $y$ are lengths of sides of the guarded space
(it holds that $x = \wrf$ if $\wrf>\bkf$ and $x = 7 - \wrf$ if
$\wrf < \bkf$, and analogously for $y$).

\noindent {\bf Critical square:} The {\em critical square} is the
square adjacent to the square of the rook in the direction of the
black king; if the rook and the black king are in the same column or
the same row, then the critical square is between them, otherwise, it
is diagonal to the square of the rook (Figure \ref{fig:room}). More
precisely, if the rook is on $(\wrf,\wrr)$ and the black king on
$(\bkf,\bkr)$, then the coordinates $(x,y)$ of the critical square are
given as follows:

$$x=\left\{ \begin{array}{ll}
\wrf,    &  \mbox{if $\wrf=\bkf$} \\
\wrf-1,  &  \mbox{if $\wrf>\bkf$} \\
\wrf+1,  &  \mbox{if $\wrf<\bkf$} \\
\end{array} \right.
\;\;\;
y=\left\{ \begin{array}{ll}
\wrr,    &  \mbox{if $\wrr=\bkr$} \\
\wrr-1,  &  \mbox{if $\wrr>\bkr$} \\
\wrr+1,  &  \mbox{if $\wrr<\bkr$} \\
\end{array} \right.
$$

\noindent {\bf Rook exposed:} The rook is {\em exposed} if white king
cannot reach it fast enough to protect it, i.e., if white (black) is
on turn and the Chebyshev distance between the rook and the white king
is greater by at least 2 (by at least 1) than the Chebyshev distance
between the rook and the black king.

\noindent {\bf Rook divides:} The white rook {\em divides} two kings
if its $x$ coordinate is (strictly) between $x$ coordinates of the two
kings, or if its $y$ coordinate is (strictly) between $y$ coordinates
of the two kings (or both).

\noindent {\bf L-pattern:} Three KRK pieces form an {\em L-pattern} if
the kings are in the same row (column), at the distance 2, and if the
rook and the white king are in the same column (row) and at the
distance 1 (Figure \ref{fig:room}).

\noindent {\bf Kings on a same edge:} The two kings are on a
  same edge if their files or ranks are both equal to 0 or 7.

\noindent {\bf Towards black king's edge move:} If the black
king is on an edge, then the white king moves towards that edge.

\paragraph{Basic Strategy}

The strategy can be defined as follows:

\begin{itemize}[leftmargin=\parindent]
 \item [{\bf 1.}] $\ImmediateMateMoveTxt$: If there is a mating move, play it;

 \item [{\bf 2.}] $\ReadyToMateMoveTxt$: If the above is not possible
   and there is a move that leads to mate in the next move, play it
   (Figure \ref{fig:room}).

 \item [{\bf 3.}] $\SqueezeMoveTxt$: If none of the above is possible,
   make a move (by the rook) that reduces the room and in the reached
   position it holds that: $(i)$ the rook is not exposed, $(ii)$ the
   rook divides the kings and $(iii)$ it is not stalemate.

 \item [{\bf 4.}] $\ApproachMoveTxt$: If none of the above is
   possible, then approach the critical square, i.e., move the king so
   the Manhattan distance between the king and the critical square is
   decreased; in the reached position, the following has to hold:
   $(i)$ the rook is not exposed, $(ii)$ the rook divides the kings or
   there is a L-pattern, $(iii)$ if the room is less than or equal to
   3, then the following holds: the white king is not on an edge and
   if its Chebyshev distance from the rook is 1, then it does not make
   a move towards the black king's edge, and $(iv)$ it is not
   stalemate. Play the approach move in a non-diagonal direction
   ($\ApproachNonDiagMoveTxt$) only if no diagonal approach move
   ($\ApproachDiagMoveTxt$) is possible.

 \item [{\bf 5.}] $\KeepRoomMoveTxt$: If none of the above is
   possible, then keep the room, i.e., move the king if that does not
   increase the Chebyshev distance from the rook; in the reached
   position, the following has to hold: $(i)$ the rook is not exposed
   and divides the kings, and $(ii)$ if the room is less than or equal
   to 3, then the following holds: the white king is not on an edge
   and if its Chebyshev distance from the rook is 1, then it does not
   make a move towards the black king's edge, and $(iii)$ it is not
   stalemate. Play the keep room move in a non-diagonal direction
   ($\KeepRoomNonDiagMoveTxt$) only if no diagonal keep room move
   ($\KeepRoomDiagMoveTxt$) is possible.

 \item [{\bf 6.}] $\RookHomeMoveTxt$: If none of the above is
   possible, then move the rook to be horizontally or vertically
   adjacent to the white king; in the reached position the following
   has to hold: $(i)$ the rook is adjacent to the black king only if
   it is guarded by the white king and $(ii)$ it is not stalemate.

 \item [{\bf 7.}] $\RookSafeMoveTxt$: If none of the above is possible, then
move the rook to some edge (other than the edge it is possibly is on);
in the reached position the following has to hold: $(i)$
either both kings are next to the rook or the Chebyshev distance
between the rook and the black king is greater\footnote{If the
  distance would be equal to 2, then the black king could approach the
  rook, so the rook would have to make the rook safe move again and
  again.} than 2, and $(ii)$ it is not stalemate.
\end{itemize}
  
On the $8 \times 8$ board, the above steps are used in the following
number of positions (in total 175168): 1512
($\ImmediateMateMoveTxt$), 4676 ($\ReadyToMateMoveTxt$), 116504
($\SqueezeMoveTxt$), 12160+4020 ($\ApproachNonDiagMoveTxt$ +
$\ApproachDiagMoveTxt$), 3160+184 ($\KeepRoomNonDiagMoveTxt$ +
$\KeepRoomDiagMoveTxt$), 32520 ($\RookHomeMoveTxt$), 432
($\RookSafeMoveTxt$).
Note that some kinds of strategy moves can be played in different ways.
For example, when the $\SqueezeMoveTxt$ move is applicable, there are
often several possibilities to play it. Although it is not necessary
for correctness (as we will show), for efficiency reasons (i.e., for
reaching mate faster) $\SqueezeMoveTxt$ should be {\em maximal} -- it
should be played so that the black king is confined to the smallest
possible room.  Also, if there are more $\RookHomeMoveTxt$ moves
applicable, the one with the smallest Manhattan distance between the
rook and the black king should be chosen. For other moves for which
there are more options, it does not matter which of them will be
selected.

For the sake of formal analysis, the above strategy differ to some
extent from Bratko's strategy, but still keeps its spirit \cite{icga}.

\section{Problem Representation}
\label{sec:representation}

The FIDE Handbook \cite{fide} is the authoritative account of the laws
of chess, but it specifies chess only informally and is suitable only
for informal reasoning. For more rigorous reasoning, the chess notions
have to be specified formally, in some strict framework. In the rest
of this section, we will discuss central issues in representing
general chess rules and notions in various computer based
frameworks. We will discuss three concrete examples -- representation
within the general purpose programming language C, within the
constraint solver URSA and within the proof assistant
\IsabelleHOL.\footnote{All formalizations, programs and proofs
  discussed in this paper are available online from
  \url{http://argo.matf.bg.ac.rs/downloads}.}

\subsection{General Chess Rules}

General chess notions can be strictly defined, for example, in terms
of the structure of natural numbers, or within Zermelo-Fraenkel set
theory (ZFC), or given axiomatically, via axioms in first order logic
(FOL), or even using a general-purpose programming language (such as C
or Haskell). For computer-supported reasoning about chess, the most
reliable approach is to have explicit definitions of chess within a
rigorous logical framework of some proof-assistant, which is usually
some variant of higher-order logic (HOL) or type theory.

Chess is a complex game, with many rules, but it turns out that there
are not many central notions that have to be used to state the
properties like correctness of strategies for chess endgames that we
are primarily concerned about. The chess game starts in an initial
position, then two players play after each other and the game proceeds
through a series of legal positions until one player wins or the game
is drawn. Transitions between positions are made by legal moves played
by the players. Therefore, to formally specify a chess game, one must
represent \emph{arbitrary chess positions}, the \emph{initial
  position}, \emph{legal positions}, \emph{legal moves}, positions
\emph{won} for a player (its opponent is checkmated) and positions
that are \emph{drawn} (it is stalemate or a checkmate cannot be
reached). Other specific definitions (e.g., the capture rules, or the
promotion rules for a pawn) are just building blocks used to define
the central notions. 
 
We will follow Hurd\cite{hurd2005} and make some important
simplifications (since we are primarily focused on endgames). First,
only pawnless games with no castling are considered. We also do not
formalize three-fold repetition of positions, and fifty-move
rule.\footnote{These two rules are not relevant for the strategy that
  we analyze: as it will be shown, it does not allow repetitions of
  positions, and it leads to a mate in 33 moves at most.} Also, the
FIDE rules state that a position is legal if it is reachable from the
initial position by a sequence of legal moves, but in the definition
of legal positions (i.e., in the $\legalposA$ predicate) we omit this
condition and the definition of the initial position. Namely, all
positions legal in the strict FIDE sense will satisfy the conditions
of our definition (there might exist some positions that satisfy our
definition, but are not legal in the strict FIDE sense). So, since we
show correctness of endgame strategies for all positions legal in a
weaker sense, our proofs will be valid also wrt.~the FIDE
definition. Although a number of specific chess rules are missing and
our theory is not a fully developed theory of the general chess, it
still precisely defines notions relevant for pawnless endgames and
gives us means to formally prove that our specific endgame definitions
are in accordance with the general chess rules.

Instead of full details of a general chess formalization
\cite{hurd2005,krk-cade}, we will give only a rough outline. Since it
needs to be strict and formal, we will present it in the style of the
proof assistant \IsabelleHOL{}. The basic notions are the
following.\footnote{Following the spirit of given representation, a
  general theory of two player strategic games (including chess) could
  be defined.}

\begin{itemize}
\item The $\sideA$ is a datatype denoting two players ($\whiteA$ and
  $\blackA$).

\item Positions are represented by a type $\pA$, characterized by the
  following components.
  \begin{itemize}
  \item A function $\onturnA: \pA \Rightarrow \sideA$ gives the player
    that is on turn in the given position.
  \item A function
    $\onsquareA: \pA \Rightarrow \squareA \Rightarrow (\sideA \times
    \pieceA)\ option$
    gives the piece on the given square in the given position (or
    returning the special value $None$ if the square is empty).  In
    the above, the datatype $\pieceA$ contains the chess pieces
    $\kingA$, $\queenA$, $\rookA$, $\bishopA$, $\knightA$ (note that
    we do not consider pawns) and the datatype $\squareA$ contains the
    squares on the board. The function implicitly ensures that there
    cannot be more than one piece on a square.
  \end{itemize}
\end{itemize}

The above notions are used in definitions of the following basic
functions.

\begin{itemize}
\item The function $\legalposA: \pA \Rightarrow \bool$ checks if the
  given position is legal.
\item The function
  $\legalmoveA: \pA \Rightarrow \pA  \Rightarrow \bool$
  checks if the second given position can be reached from the
  first given one by a legal chess move.
\item The function $\checkmateA: \pA \Rightarrow \bool$ checks if the
  player on turn is checkmated in the given position.
\end{itemize}

The above notions for chess need to meet some conditions. For example,
concrete definitions for $\legalposA$ and $\legalmoveA$ have to ensure
that legal moves can only be made between legal positions, so the
following holds:
$\legalmoveA\ \varA{p_1}\ \varA{p_2}\ \longrightarrow\ \legalposA\
\varA{p_1}\ \wedge\ \legalposA\ \varA{p_2}$
(which we can prove as a lemma).\footnote{Notice that such conditions
  hold for most (if not all) two player strategic games, so if we
  build a general theory of two player games, formulae like the given
  two would have a role of axioms.} Also, after a legal move, the
opponent is on turn,
$\legalmoveA\ \varA{p_1}\ \varA{p_2} \longrightarrow \onturnA\
\varA{p_2} = \oppA\ (\onturnA\ \varA{p_1})$
must hold (where $\oppA$ denotes the opponent side).

\paragraph{Some details of implementation in \IsabelleHOL}
For illustration, we show some auxiliary definitions from our
\IsabelleHOL{} formalization (that follows Hurd \cite{hurd2005}) that
are used in definitions of $\legalposA$, $\legalmoveA$, and
$\checkmateA$.

The $\squareA$ type is implemented as a pair of
integers\footnote{Instead of integers, natural numbers could have been
  used. However, integers allow expressing some properties using
  subtraction more easily.} -- this enables to express many chess
definitions succinctly, using arithmetic. $\legalposA$ definition must
ensure that all pieces are within the board bounds, for which the
function
$\boardA\ (\varA{f},\varA{r}) \longleftrightarrow 0 \leq \varA{f}
\wedge \varA{f} < \filesA \wedge 0 \leq \varA{r} \wedge \varA{r} <
\ranksA$
is used (global constants $\filesA = 8$ and $\ranksA = 8$, for files
and ranks, determine the size of the board). We also define functions
that for a given position $\varA{p}$ and a square $\varA{sq}$ check if
$\varA{sq}$ is empty
($\emptA\ \varA{p}\ \varA{sq} \longleftrightarrow \onsquareA\
\varA{p}\ \varA{sq} = None$),
or occupied by a piece of a given side $\varA{sd}$
($\occupiesA\ \varA{p}\ \varA{sd}\ \varA{sq} \longleftrightarrow
(\exists \varA{pc}.\ \onsquareA\ \varA{p}\ \varA{sq} = Some\
(\varA{sd}, \varA{pc}))$).

Next we define scope of each piece. For instance:

\begin{isar_code}
{\it
\begin{tabbing}
  $\rookscopeA\ (f_1,r_1)\ (f_2,r_2) \longleftrightarrow (f_1 = f_2 \vee r_1 = r_2) \wedge (f_1 \neq f_2 \vee r_1 \neq r_2)$
\end{tabbing}
}
\end{isar_code}

In a position $\varA{p}$, a square $\varA{sq_1}$ attacks $\varA{sq_2}$
if the \emph{line between them is clear}
($\clearlineA\ \varA{p}\ \varA{sq_1}\ $ $\varA{sq_2} \longleftrightarrow
(\forall \varA{sq}.\ \squarebetweenA\ \varA{sq_1}\ \varA{sq}\
\varA{sq_2} \longrightarrow \emptA\ \varA{p}\
\varA{sq})$)\footnote{Since
  squares that a knight attacks are not on the same line with the
  square that it is on, the clear line condition is always
  satisfied.}, and if there is a piece on $\varA{sq_1}$ such that
$\varA{sq_2}$ is in its scope.

\begin{isar_code}
\begin{tabbing}
  \hspace{5mm}\=\hspace{5mm}\=\hspace{5mm}\=\kill
   $\attacksA$\ $\varA{p}$\ $\varA{sq_1}$\ $\varA{sq_2}$ $\longleftrightarrow$ $\clearlineA$\ $\varA{p}$\ $\varA{sq_1}$\ $\varA{sq_2}$ $\wedge$\\
\>     ($case$\ $\onsquareA$\ $\varA{p}$\ $\varA{sq_1}$\ $of$\\
\>\>          $None$ $\Rightarrow$ False\\
\>\>       $|$ $Some$\ (\_, $\kingA$) $\Rightarrow$ $\kingscopeA$\ $\varA{sq_1}$\ $\varA{sq_2}$\\
\>\>       $|$ $Some$\ (\_, $\rookA$) $\Rightarrow$ $\rookscopeA$\ $\varA{sq_1}$\ $\varA{sq_2}$ ...)
\end{tabbing}
\end{isar_code}

A side $\varA{sd}$ is \emph{in check} in a position
$\varA{p}$ if its king is on a square $\varA{sq_1}$, and there is an
opponent's piece on some square $\varA{sq_2}$ such that it attacks the
king on $\varA{sq_1}$.

\begin{isar_code}
{\it
\begin{tabbing}
  $\incheckA$\ $\varA{sd}$\ $\varA{p}$ $\longleftrightarrow$ ($\exists$ $\varA{sq_1}$ $\varA{sq_2}$. \=$\onsquareA$\ $\varA{p}$\ $\varA{sq_1}$ = $Some$ ($\varA{sd}$, $\kingA$) $\wedge$\\
                           \>$\occupiesA$\ $\varA{p}$\ ($\oppA$\ $\varA{sd}$)\ $\varA{sq_2}$ $\wedge$ $\attacksA$\ $\varA{p}$\ $\varA{sq_2}$\ $\varA{sq_1}$)
\end{tabbing}
}
\end{isar_code}

Finally, a \emph{position is legal} (denoted by the function
$\legalposA$) if the opponent of the player on turn is not in check,
and, since we represent squares as pairs of integers, if all pieces
are within the board bounds (coordinates of their squares are between
0 and 7, which is checked by the $\boardA$ function defined
above).\footnote{It may seem that the definition of legal position
  should require that the kings are not on adjacent squares. However,
  that condition is not necessary: if the two kings are on adjacent
  squares then, following Hurd's definitions, the player who is not on
  turn would be in check (his/her king would be attacked by the other
  king), which is not legal by the above definition.}

The notions of \emph{legal move} and \emph{checkmate} (i.e., the
functions $\legalmoveA$ and $\checkmateA$) are defined in a similar
fashion. Since we will prove that our endgame strategy will always
succeed in checkmating the opponent, we do not need to formalize the
definition of a draw.

\subsection{Chess Endgames}
\label{sec:endgames}

An alternative definition of the chess game can be given for a
specific endgame, and the relevant chess rules can be described in
simpler terms, leading to an \emph{chess endgame definition}.
 
The type $\pA$ representing general chess positions can be
significantly simplified if only the positions reachable during a
specific endgame need to be considered (e.g., in a KRK endgame, only
the rules for $\kingA$ and $\rookA$ are relevant and all other pieces
and conditions describing their moves can be omitted). Simpler and
more compact representations lead to more efficient reasoning.

Additionally, the functions on the type $\pA$ that formally describe
the general chess notions (e.g., functions $\legalposA$, $\legalmoveA$
and $\checkmateA$) need not be executable (e.g., if definitions of
those functions contain quantifiers, the computer framework need not
be able to effectively compute if there is a checkmate in a given
position). However, to aid some automatic reasoning approaches, it is
desirable to have an executable representation at least for the
endgame definition (e.g., it should be possible to compute effectively
if a position is checkmate, if a move is legal, or to enumerate
effectively all legal positions satisfying some given property).

In the ideal case, both simplified and general chess representation
(and all corresponding definitions) should be described within the
same computer framework (e.g., proof assistant). In that case, some
morphism between the two representations can be defined, and the
relationship between them can be formally shown. Further, all
reasoning about the endgame properties should be done within the proof
assistant leading to highest possible reliability. Reasoning about the
endgame properties can also be done in some other systems (e.g., it
can be done by using a constraint solver or some custom-designed C
programs). In that case, the specific, simplified representation of
endgame positions and rules have to be implemented in that system, but
the relationship of such implementation with the general rules of
chess can be shown only informally, leading to a lower degree of
confidence.

In either case (a proof assistant, or some other system), the following
notions can be used to represent the endgame rules.

\begin{itemize}
\item A type $\pEG$ represents chess positions encountered during a
  specific endgame.\footnote{Note that we will use
    $\chessendgamefont{typewriter}$ font for the endgame definition
    related notions (e.g., $\pEG$ or $\pI$) and
    $\chessgeneralfont{normal}$ $\chessgeneralfont{italic}$ font for
    general chess notions (e.g., $\pA$).} It should be able to
  represent all chess positions that contain pieces relevant for the
  specific endgame (e.g., in the KRK case, two kings and the white
  rook) and all positions that can be reached from those during a play
  based on the strategy. Therefore, there can be a function
  $\abstr{\_}: \pEG \Rightarrow \pA$ that is a bijection between
  $\pEG$ and this subset of $\pA$ i.e., a function such that each
  position $\varI{p}$ from $\pEG$ represents some such position
  $\varA{p} = \abstr{\varI{p}}$ from $\pA$. The endgame domain should
  be closed under legal moves in the general domain, i.e.,
  $\forall\ \varI{p_1}, \varA{p_2}.\ \legalmoveA\ \abstr{\varI{p_1}}\
  \varA{p_2} \longrightarrow (\exists\ \varI{p_2}.\ p_2 =
  \abstr{\varI{p_2}})$.
\item The set $\chessendgameEG{\initialpositions}: \pEG\ set$ is the
  set of possible initial positions for the endgame, i.e., the set of
  all legal positions in $\pEG$ that contain all pieces relevant for
  the endgame (note that $\pEG$ can contain some positions where some
  pieces have been captured). It must hold that this set represents
  exactly all legal positions from $\pA$ that contain all relevant
  pieces (correctness of the endgame strategy is usually formulated
  for all plays that start in a position from this set).
\item The function
  $\chessendgameEG{\legalpos}: \pEG \Rightarrow \bool$ checks if the
  given position is legal. For any position $\varI{p}$ of type $\pEG$
  it must hold that
  $\chessendgameEG{\legalpos}\ \varI{p}\ \longleftrightarrow\
  \legalposA\ \abstr{\varI{p}}$.
\item The function
  $\chessendgameEG{\legalmove}: \pEG \Rightarrow \pEG \Rightarrow
  \bool$
  checks if the second given position can be reached from the first
  given position by a legal move. It must hold that for any given
  positions $\varI{p_1}$ and $\varI{p_2}$ of type $\pEG$ it holds that
  $\chessendgameEG{\legalmove}\ \varI{p_1}\ \varI{p_2}\
  \longleftrightarrow\ \legalmoveA\ \abstr{\varI{p_1}}\
  \abstr{\varI{p_2}}$.
\item The function
  $\chessendgameEG{\checkmate}: \pEG \Rightarrow \bool$ checks
  if the player on turn is checkmated in a given position. It must
  hold that for any position $\varI{p}$ of type $\pI$ it holds that
  $\chessendgameEG{\checkmate}\ \varI{p} \longleftrightarrow
  \checkmateA\ \abstr{\varI{p}}$.
\end{itemize}

The above conditions are necessary (and sufficient) to prove in order
to show that if we prove some property for the endgame then that
property holds wrt. the general chess rules, too.

\subsubsection{A Case Study of KRK}

We will present several instances of the KRK endgame definition. 
The relevant pieces are the two kings and the white rook (that could 
be captured and our representation needs to cover that situation, too).

\paragraph{\IsabelleHOL{} and Records}

The most natural way of representing the position is to pack all
relevant information into a record (a structure) or an array. In
\IsabelleHOL{} we define the following record.

\begin{isar_code}
{\tt
\begin{tabbing}
  \hspace{3mm}\=\hspace{3mm}\=\kill
{\bf record} $\KRKPosition$ = \\
\> $\WK{}$ :: "$\squareA$" (* position of white king *)\\
\> $\BK{}$ :: "$\squareA$" (* position of black king *)\\
\> $\WRopt{}$ :: "$\squareA$ option" (* position of white rook ($None$ if captured) *)\\
\> $\WhiteOnTurn{}$ :: "$bool$" (* Is white on turn? *)
\end{tabbing}
}
\end{isar_code}

In order to represent a chessboard position, such record has to
meet several conditions. First, the following condition checks
whether all pieces are within the board bounds:

\begin{isar_code}
{\tt
\begin{tabbing}
$\boardA$  ($\WK{\varI{p}}$) $\wedge$ $\boardA$ ($\BK{\varI{p}}$) $\wedge$ ($\neg
\WRcaptured{\varI{p}}$ $\longrightarrow$ $\boardA$ ($\WR{\varI{p}}$))
\end{tabbing}
}
\end{isar_code}

$\WRcaptured{\varI{p}}$ denotes that the rook is captured in position
$\varI{p}$, i.e., that $\WRopt{p} = None$. The following condition
checks whether pieces are on different squares

\begin{isar_code}
{\tt
\begin{tabbing}
  $\WK{\varI{p}}$ $\neq$ $\BK{\varI{p}}$ $\wedge$
  ($\neg \WRcaptured{\varI{p}}$ $\longrightarrow$ $\WR{\varI{p}}$
  $\neq$ $\WK{\varI{p}}$ $\wedge$ $\WR{\varI{p}}$ $\neq$
  $\BK{\varI{p}}$)
\end{tabbing}
}
\end{isar_code}

Note that this definition uses the notion of $\squareA$, and the
$\boardA$ predicate which are also used in the definition of the
general chess rules. Since both the general chess definition and the
endgame definition are given in the same system (proof
assistant), we could reuse such definitions.

$\pI$ is the type consisting of all $\KRKPosition$ records that
satisfy the two conditions given above. The abstraction function
($\abstr{\_}$) that maps KRK positions to general chess positions that
they represent is defined as follows:

\begin{isar_code}
{\tt
\begin{tabbing}
  $\onsquareA$\ $\varI{p}$ = $\lambda$\ $\varI{sq}$.\ (\=if $\WK{\varI{p}}$ = $\varI{sq}$ then $Some\ (\whiteA, \kingA)$ \\
  \>     else if $\BK{\varI{p}}$ = $\varI{sq}$ then $Some\ (\blackA, \kingA)$\\
  \>     else if $\WRopt{\varI{p}}$ = Some $\varI{sq}$ then $Some\ (\whiteA, \rookA)$\\
  \>     else $None$)"
\end{tabbing}
}
\end{isar_code}

Auxiliary functions that lead to the $\legalposI$ and $\legalmoveI$
definitions are reformulated for the endgame. The following function
checks if the black king is attacked, and is used in the definition of
checkmate, along with the $\legalposI$ and $\BKCannotMove$ definitions
(that are not shown here, but are available in the formal proof
documents).

\begin{isar_code}
{\tt
\begin{tabbing}
\hspace{5mm}\=\hspace{5mm}\=\kill
$\WRattacksBK$\ $\varI{p}$ $\longleftrightarrow$\\
\> $\neg$ $\WRcaptured{\varI{p}}$ $\wedge$ $\rookscopeA$\ ($\WR{\varI{p}}$)\ ($\BK{\varI{p}}$)\ $\wedge$ $\neg$ $\squarebetweenhvA$\ ($\WR{\varI{p}}$)\ ($\WK{\varI{p}}$)\ ($\BK{\varI{p}}$)\\[2mm]
$\checkmateI$\ $\varI{p}$\ $\longleftrightarrow$\ $\legalposI\ \varI{p}$ $\wedge$ $\BKCannotMove\ \varI{p}$ $\wedge$ $\WRattacksBK$\ $\varI{p}$
\end{tabbing}
}
\end{isar_code}

The $\rookscopeA$ definition is taken from the general chess
formalization, but the notion of the black king being attacked is
specific to KRK ($\squarebetweenhvA$ call checks only if the white
kings blocks the line between the rook and the black king horizontally
or vertically) and such simple way is not correct wrt.~the general
chess rules (in general chess other pieces must be taken into
account). Because of such differences, it is essential to have a
formal link between the two layers.

\paragraph{C and Structures}

In C, it is natural to represent squares by two integer
coordinates. Also, there is no built-in option type in C so, for
simplicity, we just add a flag that tells if the rook has been
captured (if the rook is captured then its two coordinates become
irrelevant). The type $\pI$ is the following:

{\footnotesize
\begin{verbatim}
typedef struct pos {
  bool bWhiteOnTurn;
  bool bRookCaptured;
  unsigned char WKx, WKy, WRx, WRy, BKx, BKy;
} KRKPosition;
\end{verbatim}
}

For efficient storing, we represent each position by a bitvector of
length 20: each of the three pieces by two triples of bits (as a
triple of bits gives 8 possible values, corresponding to the default
chessboard size) and two bits for representing which player is on turn
and whether the rook has been captured. For instance, the position
shown left in Figure~\ref{fig:room}, with white on turn, is
represented by the numbers $(3,2)-(5,4)-(2,6)-1-0$, i.e,. by the
following bitvector: $01101010110001011010$.  We implemented functions
for transforming the above structure into bitvectors and back.

Unlike \IsabelleHOL, where the conditions that the record must satisfy
are explicit, in C these conditions are ensured by additional
functions.

It is easy to formulate all relevant chess notions and rules. For
instance, the definition of mate is formulated in C as
follows.\footnote{Presented specifications in different frameworks are
  equivalent and substantially the same. However, there are still some
  minor differences (e.g., in naming conventions, in grouping of some
  conditions into predicates, etc), just as there are different
  programming styles.}

{\footnotesize
\begin{verbatim}
bool Mate(KRKPosition p) {
    return LegalPositionBlackToMove(p) && BlackCannotMove(p) && WRAttacksBK(p);
}
\end{verbatim}
}

\paragraph{URSA Constraint Solver, Bitvectors, and SAT}
The URSA constraint solver is based on bitvectors and reduction
(``bit-blasting'') to SAT. In URSA, a position can be conveniently and
naturally specified by six triplets of bits (two for each of the three
pieces) plus bits for representing which player is on turn and whether
the rook has been captured (as in the C version).  Therefore, in URSA
we represent positions with 20-bit numbers, and we developed
procedures for transformation from individual pieces of information to
20-bit numbers and back.

Since the URSA language is C-like, it is easy to formulate all
relevant chess notions and rules, very similarly as in the C version.
For instance:

{\footnotesize
\begin{verbatim}
procedure Mate(nPos, bMate) {
  call LegalPositionBlackToMove(nPos, bLegalPositionBlackToMove);
  call BKCannotMove(nPos, bBKCannotMove);
  call WRAttacksBK(nPos, bBKAttacked);
  bMate = bLegalPositionBlackToMove && bBKCannotMove && bBKAttacked;
}
\end{verbatim}
}

Once the URSA specification is made, by merits of the URSA
  system, we can immediately get a representation of properties of the
  KRK endgame in the language of SAT -- bitvectors are vectors of
  Boolean variables and each URSA procedure call generates a Boolean
  formula constraining the parameters.

\paragraph{Linear Integer Arithmetic (LIA)}

Another possibility is to represent KRK positions and all relevant
predicates using the language of linear integer arithmetic (LIA). The
type $\pI$ then consists of integers $\varI{\wkf}$, $\varI{\wkr}$,
$\varI{\bkf}$, $\varI{\bkr}$, $\varI{\wrf}$, and $\varI{\wrr}$, the
Boolean $\varI{WhiteOnTurn}$, and the Boolean $\varI{WRCaptured}$. We
constructed such specification in terms of LIA within the
\IsabelleHOL{} proof assistant. Here is an example definition.

\begin{isar_code}
\begin{tabbing}
  \hspace{5mm}\=\hspace{5mm}\=\kill
$\LIArookscope$\ $\varI{f_1}$\ $\varI{r_1}$\ $\varI{f_2}$\ $\varI{r_2}$\ $\longleftrightarrow$\ $(\varI{f_1} = \varI{f_2}\ \vee\ \varI{r_1} = \varI{r_2})\ \wedge\ (\varI{f_1} \neq \varI{f_2}\ \vee\ \varI{r_1} \neq \varI{r_2})$
\end{tabbing}
\end{isar_code}

Note that these definitions are very similar to the ones based on
records and the option type, but they use only LIA constructs.
Because of that, the \IsabelleHOL{} system can automatically transform
such definitions to the SMT-LIB input format and apply SMT solvers,
which is the main method that we will use for reasoning in
\IsabelleHOL.

\subsection{Chess Endgame Strategies}
\label{sec:endgamestrategies}

Given the representation of the chess (endgame) rules, endgame
strategies can be represented. Without loss of generality, we can only
consider strategies for the white player. We can think about a
strategy as a function that maps a given position to a position that
is going to be reached after playing a strategy move. This function
does not need to be total (w.r.t.~the set of all positions with the
relevant pieces on the board), for example, it need not be defined for
positions that cannot be reached during the endgame. Instead of
functions, sometimes a better choice could be to allow
non-deterministic strategies and to model strategies by
relations. Namely, non-deterministic strategies can be underspecified
and can have much simpler definitions than corresponding deterministic
functions -- relations should describe only those aspects that are
necessary for correctness, while aspects related to efficiency could
be omitted from the specification and postponed (see Section
\ref{subsec:citius}).  Therefore, we have the following choices for
the strategy definition.

\begin{itemize}
\item A relation $\rel{\wsA}: \pA \Rightarrow \pA \Rightarrow \bool$.
  There can be more than one position reachable by a strategy from a
  given position in one ply. The strategy must be \emph{legal} i.e.,
  it must give only legal moves:

$$\rel{\wsA}\ \varA{p}_1\ \varA{p}_2  \Rightarrow \legalmoveA\ \varA{p}_1\ \varA{p}_2$$

\item A deterministic-function $\dfun{\wsA}: \pA \Rightarrow \pA\ opt$
  ($opt$ denotes the option type, like in \IsabelleHOL).  This
  function corresponds to one specific instance of the strategy and
  returns the (single) move that white following the strategy
  should play in the given position (or the special value
  $\mathit{None}$, if the strategy is undefined for the given
  position).
\end{itemize}

If both the relation $\rel{\wsA}$ and the function $\dfun{\wsA}$ are
defined, then it is natural to require that they agree:
  $$\dfun{\wsA}\ \varA{p} = \varA{p'} \longrightarrow  \rel{\wsA}\ \varA{p}\ \varA{p'}.$$

We also consider a non-deterministic function
$\ndfun{\wsA}: \pA \Rightarrow \pA\ set$, defined as
$\ndfun{\wsA}\ \varA{p} = \{\varA{p'}\ |\ \rel{\wsA}\ \varA{p}\
\varA{p'}\}.$
Note that given the implementation of $\rel{\wsA}$, it can still be
non-trivial to obtain an implementation of $\ndfun{\wsA}$.

We will also consider the following relation:

\begin{itemize}
\item $\rel{\bA}: \pA \Rightarrow \pA \Rightarrow \bool$ is a relation
  such that $\rel{\bA}\ \varA{p}\ \varA{p'}$ holds whenever a position
  $\varA{p'}$ can be reached from the position $\varA{p}$ by a legal
  move of black. Note that this is just a restriction
  of the relation $\legalmoveA$ on the set of positions in which black
  is on turn.
\end{itemize}

Since the play of black is not restricted, we can't consider
$\dfun{\bA}$ that would be analogous to $\dfun{\wsA}$.
  
Although we assume strategies on the general chess position type
$\pA$, it suffices to define them only on the endgame type $\pI$
(e.g., we consider the relation
$\rel{\wsI}: \pI \Rightarrow \pI \Rightarrow \bool$).  Every strategy
definition on the type $\pI$ can naturally be lifted and yields a
strategy on the type $\pA$. Namely, given a strategy relation
$\rel{\wsI}$ defined in the endgame terms, two positions $\varA{p_1}$
and $\varA{p_2}$ of the type $\pA$ are connected by the strategy
relation $\rel{\wsA}$ defined in the general chess terms, i.e.,
$\rel{\wsA}\ \varA{p_1}\ \varA{p_2}$ if and only if both can be
represented by the type $\pI$ i.e., if there are positions
$\varI{p_1}$ and $\varI{p_2}$ of the type $\pI$ such that
$\abstr{\varI{p_1}} = \varA{p_1}$, $\abstr{\varI{p_2}} = \varA{p_2}$
and $\rel{\wsI}\ \varI{p_1}\ \varI{p_2}$ holds. If a strategy function
is lifted, then it returns $None$ for all positions that cannot be
represented by the type $\pI$.  Again, this is important for
maintaining the link with the general chess game.

\subsubsection{A Case Study of the Strategy for KRK}
\label{sec:krkstrategy}
We have developed several implementations of the KRK endgame strategy
described in Section \ref{sec:ourbratko}, and then proved its
correctness. Each implementation relies on some previously described
KRK endgame representation.

\paragraph{Defining strategy conditions}
For illustration, we show some auxiliary definitions that lead to the
strategy definition. For example, in \IsabelleHOL{} the function that
checks if a position $\varI{p'}$ can be reached from a position
$\varI{p}$ by an $\ImmediateMateMoveTxt$ move is formalized as
follows (assuming that auxiliary predicates $\BKCannotMove$ and
$\WRattacksBK$ were previously defined).

\begin{isar_code}
\begin{tabbing}
  \hspace{5mm}\=\kill
  $\ImmediateMateCond\ \varI{p}\ \varI{p'} \longleftrightarrow \BKCannotMove\ \varI{p'} \wedge \WRattacksBK\ \varI{p'}$
\end{tabbing}
\end{isar_code}

Similarly, the $\RookHomeMoveTxt$ condition is formalized as follows
(the divide attempt requires that the white rook is in a file or rank
next to the white king):

\begin{isar_code}
\begin{tabbing}
\hspace{5mm}\=\kill
$\RookHomeCond \ \varI{p}\ \varI{p'} \longleftrightarrow$\\
\>     $\DivideAttempt\ \varI{p'}\ \wedge$\\
\>     $(\kingscopeA\ (\BK{\varI{p'}})\ (\WR{\varI{p'}}) \longrightarrow \kingscopeA\ (\WK{p'})\ (\WR{p'}))\ \wedge$\\
\>   $(\BKCannotMove\ \varI{p'}\ \longrightarrow\ \WRattacksBK\ \varI{p'})$
\end{tabbing}
\end{isar_code}

The corresponding definition in URSA is very similar\footnote{
The condition that the position is not stalemate makes one of 
small differences between the URSA and Isabelle/HOL specifications.
In Isabelle/HOL, repeating the constraint on stalemate in conditions
for all move kinds would give very large formulae, so in the 
definition of the strategy relation that condition is factored out 
and included only once, globally. On the other hand, URSA implements 
subformula sharing, so no significant overhead is incurred if a 
constraint is repeated several times.}

{\footnotesize
\begin{verbatim}
procedure RookHomeCond(nPos1, nPos2, bRookHomeCond) {
  call LegalMoveWR(nPos1, nPos2, bLegalMoveWR);
  call DivideAttempt(nPos1, nPos2, bDivideAttempt);
  call BKNextWR(nPos2, bBKNextWR);
  call WKNextWR(nPos2, bWKNextWR);
  call Stalemate(nPos2, bStalemate);
  bRookHomeCond = bLegalMoveWR && bDivideAttempt && (!bBKNextWR || bWKNextWR) && !bStalemate;
}
\end{verbatim}
} 

The corresponding definitions in C and LIA are also very similar.

\paragraph{Defining the strategy relation}
The applied move must be the first one whose condition
holds. Therefore, for each move we must have a function that checks if
that move links the two given positions, and a function that checks if
the strategy move is not applicable in a given position. Notice that
these two are not just opposites of each other, since the latter
requires rejecting all possibilities for this strategy move to be
played from the given position. In \IsabelleHOL{}, we introduce the
function $\kingssquare\ (f, r)\ k$ that for an index $k$ between 1 and
8, gives coordinates of 8 squares that surround the given central
square $(f, r)$. Similarly, the function $\rookssquare\ (f, r)\ k$ for
$k$ between 1 and 16 gives all squares that are in line with the rook
(first horizontally, and then vertically). Combined, these give the
enumeration of all possible moves of white (indices from 1 to 8
correspond to the king moves, and from 9 to 24 to the rook moves). We
show this only for $\ImmediateMateMoveTxt$ in \IsabelleHOL, as other
moves follow a similar pattern.

Since we want to have all our definitions executable and we want to
deal only with quantifier-free SMT formulae, we must introduce bounded
quantification (that is unfolded into a finite conjunction). Then we
can define predicates that encode that a certain kind of move cannot
be applied.

\begin{isar_code}
{\tt
\begin{tabbing}
\hspace{5mm}\=\kill
$all\_n\ P\ n$ $\longleftrightarrow$ $\forall$ $i$. $1 \le i \wedge i \le n$ $\longrightarrow$ $P\ i$\\[1mm]
$\noImmediateMateWK$\ $\varI{p}$ $\longleftrightarrow$ $all\_n$ 8 ($\lambda$ $k$. let $\varI{sq}$ = $\kingssquare$ $(\WK{\varI{p}})$ $k$ in \\
\> $\WKcanmoveto$ $\varI{p}$ $\varI{sq}$ $\longrightarrow$ $\neg$ $\ImmediateMateCond$ $\varI{p}$ $(\moveWK\ \varI{p}\ \varI{sq}))$\\[1mm]
$\noImmediateMateWR$ $\varI{p}$ $\longleftrightarrow$ $all\_n$ 16 ($\lambda$ $k$. let $\varI{sq}$ = $\rookssquare$ ($\WR{\varI{p}}$) $k$ in \\
\> $\WRcanmoveto$ $\varI{p}$ $\varI{sq}$ $\longrightarrow$ $\neg$ $\ImmediateMateCond$ $\varI{p}$ $(\moveWR\ \varI{p}\ \varI{sq}))$\\[1mm]
$\noImmediateMate$ $\varI{p}$ $\longleftrightarrow$ $\noImmediateMateWK$ $\varI{p}$ $\wedge$ $\noImmediateMateWR$ $\varI{p}$
\end{tabbing}
}
\end{isar_code}

Note that the mating move can be performed only by the rook, and we
have formally proved that in Isabelle/HOL, so the search for a mating
move does not need to consider the moves of the king.

Since URSA, C and the quantifier-free fragment of LIA do not support
quantifiers, conditions like the above are expressed by a finite
conjunction or a loop.

Finally, we can introduce the relation
$\StrategyWhiteMove\ \varI{p}\ \varI{p'}\ m$, encoding that a position
$\varI{p'}$ is reached from a position $\varI{p}$ after a strategy
move of a kind $m$. We show only a fragment of the definition in
\IsabelleHOL{} (the C, URSA and LIA definitions are very similar).

\begin{isar_code}
{\tt
\begin{tabbing}
\hspace{5mm}\=\kill
$\MoveKind = \ImmediateMateMove\ |\ \ReadyToMateMove\ |\ \SqueezeMove\ |\ \ApproachDiagMove\ |$\\
\>  $\ApproachNonDiagMove\ |\ \KeepRoomDiagMove\ |\ \KeepRoomNonDiagMove\ |\ \RookHomeMove\ |\ \RookSafeMove$\\[2mm]
\hspace{5mm}\=\hspace{5mm}\=\hspace{5mm}\=\hspace{5mm}\=\hspace{5mm}\=\hspace{5mm}\=\hspace{5mm}\=\hspace{5mm}\=\hspace{5mm}\=\hspace{5mm}\=\kill
$\StrategyWhiteMove$\ $\varI{p}$\ $\varI{p'}$\ $\varI{m}$ $\longleftrightarrow$\\
  (if $\varI{m}$ = $\ImmediateMateMove$ then\\
\>      $\legalmoveWR$ $\varI{p}$ $\varI{p'}$ $\wedge$ $\ImmediateMateCond$ $p$ $\varI{p'}$\\
   else \\
\>      $\noImmediateMate$ $\varI{p}$ $\wedge$ \\
\>      if $\varI{m}$ = $\ReadyToMateMove$  then\\
\>\>         $\legalmoveWhite$ $\varI{p}$ $\varI{p'}$ $\wedge$ $\ReadyToMateCond$ $\varI{p}$ $\varI{p'}$\\
\>      else \\
\>\>         $\noReadyToMate$ $\varI{p}$ $\wedge$ \\
\>\>         \ldots\\
\>\>\>\>\>\>\>\>                           if $\varI{m}$ = $\RookSafeMove$ then\\
\>\>\>\>\>\>\>\>\>                              $\legalmoveWR$ $\varI{p}$ $\varI{p'}$ $\wedge$ $\RookSafeCond$ $\varI{p}$ $\varI{p'}$\\
\>\>\>\>\>\>\>\>                           else $\varI{False}$)
\end{tabbing}
}
\end{isar_code}

Note that this is our strategy relation $\rel{\wsI}$, but it is
parametrized by a move type (therefore, we can consider the relation
${\rel{\wsI}^\varI{m}}$, where $m$ is $\ImmediateMateMoveTxt$,
$\ReadyToMateMoveTxt$, etc.).

\paragraph{Defining the strategy function}
Although the strategy relation permits to play several different moves
in some
position, the ultimate goal is usually to reduce that to an
executable function that calculates a single white player move for
each position when it is on turn. 

Defining deterministic strategy function $\dfun{\wsI}$ requires a bit
more effort. A function can iterate through all legal moves of white
pieces until it finds a first move that satisfies the relational
specification. An interesting exception is the $\SqueezeMoveTxt$
move. To make the strategy more efficient, the maximal
$\SqueezeMoveTxt$ (the one that confines the black king the most) is
always played (if there are several such moves, the first one found in
the iterating process is used).

The strategy function definition in \IsabelleHOL{} uses several
auxiliary functionals. The functional
$\firstlegalWK\ \varI{p}\ \varI{cond}$ takes a position $\varI{p}$ and
a condition $\varI{cond}$ (a unary predicate formulated on the set of
all positions), iterates through all possible moves of the white king
(using the function $\kingssquare{}$), and returns the index (a number
between 1 and 8) of the first legal move (i.e., the move with the
minimal index) that leads to a position that satisfies the given
condition $\varI{cond}$, or zero if there is no such move. The
functionals $\firstlegalWR$ and $\firstlegalWhite$ are defined
similarly.  For example, the value of the expression
$\firstlegalWR{}\ \varI{p}\ \ImmediateMateCond$ is the index of the
first mating move by the white rook starting from the position
$\varI{p}$, or zero if the black cannot be mated in one move.  We also
introduce the functional
$\minlegalWR\ \varI{p}\ \varI{cond}\ \varI{score}$, that takes a
position $\varI{p}$, a condition $\varI{cond}$ and a function that
assigns penalty scores to positions. The functional $\minlegalWR$
returns the index of the move of the white rook (a number between 1
and 16) that leads into the position that has the minimal penalty
score among all such positions that satisfy the given condition
$\varI{cond}$ (if there is more than one such move, the first one
i.e., the minimal index is returned).

Using these auxiliary functions, the strategy
function is defined in \IsabelleHOL{} as follows.

\begin{isar_code}
{\tt
\begin{tabbing}
\hspace{5mm}\=\kill
 "$\StrategyWhiteMoveFun$ $\varI{p}$ = \\
\>   (l\=et i = $\firstlegalWR$\ $\varI{p}$ $\ImmediateMateCond$\\
\>     \>in if i > 0 then ($\movewhite$ $\varI{p}$ (i+8), $\ImmediateMateMove$) \\
\>    else l\=et i = $\firstlegalWhite$\ $\varI{p}$ $\ReadyToMateCond$ \\
\>     \>in if i > 0 then ($\movewhite$ $\varI{p}$ i, $\ReadyToMateMove$) \\
\>    else l\=et i = $\minlegalWR$\ $\varI{p}$\ ($\SqueezeCond$\ $\varI{p}$)\ $\room$ \\
\>     \>in if i > 0 then ($\movewhite$ $\varI{p}$ (i+8), $\SqueezeMove$) \\
\>...
\end{tabbing}
}
\end{isar_code}

This definition always gives the $\SqueezeMoveTxt$ that maximally
reduces the room.

This definition can be executed from within Isabelle/HOL (by means of
\texttt{value} command) or its code can be exported in one of the
supported functional languages (e.g., Haskell).

In URSA, the function is implemented similarly. A loop through all
possible move indices is used to find the one that satisfies the
current move condition.

\section{Reasoning Methods}
\label{sec:reasoning}

In this section we will describe several approaches for proving
correctness of a given chess endgame strategy, i.e., for proving that
the strategy for white is winning starting from any of relevant legal
positions.  We define that $\rel{\wsA}$ is a \emph{winning strategy}
for white on a set of positions $\varA{I}$ with white on turn if all
positions in $\varA{I}$ are $\wsA$-\emph{winning positions} for
white. A position is $\wsA$-winning for white if each play starting
from it terminates in a position where black is checkmated, given that
white follows the strategy.\footnote{Note that every $WS$-winning
  position is a winning position (assuming perfect play), but the
  opposite does not necessarily hold.} More formally, $\wsA$-winning
positions can be defined inductively:
{\em (i)} A position is $\wsA$-winning, if white, following the strategy
  $\wsA$, immediately mates;
{\em (ii)} A position is $\wsA$-winning if each strategy move by white
  followed by any legal move of black leads into a $\wsA$-winning
  position.

It is suitable, and sometimes even necessary to use computer support
with chess rules and the strategy explicitly defined within a strict
environment (proof assistant, theorem prover, constraint solving
system, programming language etc.). Some of these proving approaches
require that the implementation of the strategy is executable (i.e.,
that functions $\rel{\wsI}$, $\ndfun{\wsI}$, or $\dfun{\wsI}$ are
implemented and used). We will assume that the reasoning will be
performed on the endgame level, while the link to the general chess
rules should be ensured as discussed previously. Two approaches can be
used.

\begin{description}
\item[Exhaustive retrograde analysis.] This approach assumes that the
  strategy is represented by a lookup table (endgame database) that
  assigns a strategy move to each relevant position. Then, using a
  retrograde procedure (in the style of Thompson's work
  \cite{thompson-lookup}), it is verified that the endgame lookup
  table ensures win for white. If the strategy is represented
  algorithmically (e.g., by the function $\dfun{\wsI}$), then the
  strategy move for each position is computed and stored into the
  lookup table. This approach is straightforward, but it does not
  provide a high-level, understandable and intuitive, argument on {\em
    why} the strategy really works.

\item[High-level conjectures.] Within this approach, correctness of
  the strategy relies on several conjectures (e.g., invariants for
  various strategy moves, termination conditions) which, when glued
  together, imply that the strategy is winning. Conjectures can be
  proved either by enumerating all possibilities or by some more
  sophisticated reasoning methods, either manually or by using
  computer support. In the latter case, the conjectures can be proved
  either formally, within a proof assistant, or informally checked
  using a general purpose programming language or a constraint solver.
\end{description}

\subsection{Retrograde Analysis}
\label{subsec:retrograde}

In this section we present a retrograde-style, enumeration-based
procedure that can be used for showing correctness of a
strategy\footnote{A slightly modified algorithm can be used for
  computing look-up table for optimal play, as Thompson did
  \cite{thompson-lookup}.}.

Let $\varA{I}$ denote a set of all initial positions for which we
claim that every play (with white following the strategy) starting
from them will terminate with black checkmated. To apply the
procedure, the set $\varA{I}$ must be closed under strategy moves of
white followed by arbitrary legal moves of black, so every strategy
play that starts from a position in $\varA{I}$ always remains in
$\varA{I}$. In the case of KRK we will assume that $\varA{I}$ equals
the set $\chessendgame{\initialpositions}$ defined in Section
\ref{sec:endgames} as a set of all legal positions with only the three
relevant pieces on the board (for the strategy that we consider, the
condition on $\varA{I}$ is met, which can be easily
proved). Correctness of the strategy can be proved by showing that
each position from $\varA{I}$ is a winning position.

WS-winning positions could be calculated by using a direct recursion,
but it is much better to apply dynamic programming. For simplicity, we
will assume that a deterministic strategy $\dfun{\chessgeneral{\ws}}$
is given, and defined for all positions $\varA{p}$ reachable from
$\varA{I}$ and that functions $\chessgeneral{\checkmate}$,
$\ndfun{\chessgeneral{\bb}}$, and $\dfun{\chessgeneral{\ws}}$ are
executable. At the beginning of the procedure it is checked that
$\varA{I}$ is closed in the above sense, i.e., that
$\ndfun{\chessgeneral{\bb}}\ (\dfun{\chessgeneral{\ws}}\ \varA{I})
\subseteq I$.
The set $\varA{\winning}$ will contain positions determined to be
winning. It will be initialized to all positions from which white
following the strategy immediately mates, and those positions will be
removed from the set $\varA{I}$. After that, the following is
repeated. Among the remaining positions (which are not yet in the set
$\varA{\winning}$), the set $\varA{S}$ is found, containing all
positions $p$ such that: after a strategy move by white from $p$ to
$p'$, it is not stalemate and all possible moves of black in $p'$ lead
to positions already in $\varA{\winning}$. These are also the winning
positions and we transfer them from the set $\varA{I}$ to the set
$\varA{\winning}$. The process terminates if all positions are
determined to be winning (in that case, the set of remaining positions
$\varA{I}$ is empty), or if there is no change made by the current
iteration (in that case, the set $\varA{S}$ is empty). This procedure
can be implemented within a function
$\chessgeneral{\retrograde}: \chessgeneral{P}\ set \Rightarrow \bool$:

{\tt
\begin{tabbing}
  \hspace{5mm}\=\hspace{5mm}\=\kill
  function $\chessgeneral{\retrograde}(\varA{I})$\\
  begin\\
  \>$\varA{S} := \{\varA{p} \in \varA{I}\ |\ \chessgeneral{\checkmate}\ (\dfun{\chessgeneral{\ws}}\ \varA{p})\}$\\
  \>$\varA{\winning} := \varA{S},\quad \varA{I} := \varA{I} \setminus \varA{S}$\\
  \>repeat\\
  \>\>$\varA{S} = \{\varA{p} \in \varA{I}\ |\ \ndfun{\chessgeneral{\bb}}\ (\dfun{\chessgeneral{\ws}}\ \varA{p}) \neq \emptyset \,\wedge\, \ndfun{\chessgeneral{\bb}}\ (\dfun{\chessgeneral{\ws}}\ \varA{p}) \subseteq \varA{\winning}\}$\\
  \>\> $\varA{\winning} := \varA{\winning} \cup \varA{S}, \quad \varA{I} := \varA{I} \setminus \varA{S}$\\
  \>until $\varA{I} = \emptyset\ \vee\ \varA{S} = \emptyset$ \\
  \>return $\varA{I} = \emptyset$\\
  end
\end{tabbing}
}

The procedure runs in a BFS fashion and positions are added to the set
$\varA{\winning}$ in increasing order of the number of moves needed to
checkmate black. The central loop invariant is that after $k$
iterations of the loop the set $\varA{\winning}$ contains all
positions from $\varA{I}$ for which white that follows the strategy
mates the black is mated after at most $k$ moves of the black.  From
that, it can be easily proved that the strategy
$\dfun{\chessgeneral{\ws}}$ is winning strategy on the set $\varA{I}$
iff the function $\chessgeneral{\retrograde}$ returns {\tt true}.

The procedure can also provide the longest possible game length, given
that white follows the strategy.

Note that the analysis could be easily modified to use the
non-deterministic definition of the strategy
$\ndfun{\chessgeneral{\ws}}$ instead of the deterministic version
$\dfun{\chessgeneral{\ws}}$.

Note that although we have defined the function $\retrograde$ in the
general chess terms, it can be implemented in chess-endgame terms.

\subsubsection{A Case Study of the Strategy for KRK}
\label{subsubsec:retrograde-discussion}

Given the strategy implementation, it was rather straightforward to
implement the above function in \IsabelleHOL{} and in C (the relevant
part of the code was only around a hundred lines of code and took only
half a day to write).

The retrograde analysis revealed some bugs in the initial
implementation, and once they were fixed, confirmed that the strategy
is correct. Therefore, this approach proved to be very suitable for
rapid detection of bugs in the strategy implementation, without going
into any deeper analysis of its properties.
  
There are 175\,168 legal KRK positions with the three pieces on board
and white on turn. It turns out that white always reaches win within
33 moves (i.e., within 65 plies). Due to the large number of positions
and plies, this approach is hardly applicable without computer
support. The check of correctness of the given KRK strategy using the
C program is done in around 5s.\footnote{All running times are
  obtained on a cluster with 32 dual core 2GHz Intel Xeon processors
  with 2GB RAM per processor. All the tests were run sequentially, and
  no parallelism was employed. The URSA system was used with its
  default SAT solver -- clasp
  (\url{http://www.cs.uni-potsdam.de/clasp/}). Although the solving
  process is deterministic, the running times can vary to some extent
  (no more than 10\%), but since the exact information on time spent
  is not critical, we kept the experiment simple and performed all
  measurements only once.}

\subsection{High-Level Conjectures}

Correctness of a strategy can be proved using a more abstract
approach. The central statement can rely on a number of auxiliary,
high-level conjectures (lemmas) that combined together lead to the
correctness arguments (that the strategy $\rel{\wsA}$, i.e.~its
counterpart $\rel{\chessendgameEG{\ws}}$, is winning), but also
provide insights into why the strategy really works. Such auxiliary
conjectures can be proved in different ways. In the following we will
focus on the KRK strategy $\rel{\wsI}$, but the method can be easily
applied to other strategies.

Many properties of the strategy can be formulated by lemmas of the
following form ($\varI{p}_i$ are positions with white on turn,
$\varI{p}_i'$ are positions with black on turn, and $\varI{m}_i$ are move
types, e.g., $\ImmediateMateMoveTxt$; a notation 
$\underset{i\in\{0..k\}}{\mathlarger{\forall}}\varI{p}_i\,\varI{m}_i\,\varI{p}_i'$
is just a shorthand for
$\forall \varI{p}_0\,\varI{m}_0\,\varI{p}_0'\,\varI{p}_1\,\ldots
  \,\varI{p}_k\,\varI{m}_k\,\varI{p}_k'\,\varI{p}_{k+1}$):

$$
\underset{i\in\{0..k\}}{\mathlarger{\forall}}\varI{p}_i\,\varI{m}_i\,\varI{p}_i'.\quad \Pre\ \varI{p}_0\ \wedge\ \underset{i\in\{0..k\}}{\Seq}\ \varI{p}_i\ \varI{m}_i\ \varI{p}_i'\ \varI{p}_{i+1} \longrightarrow \underset{i\in\{0..k\}}{\Post}\ \varI{p}_i\ \varI{m}_i\ \varI{p'}_i
$$

The predicate $\Pre$ denotes preconditions for $\varI{p}_0$.  For
example, it could express that a position $\varI{p}_0$ is legal, or
that all relevant pieces are on the board, but it could also give some
additional constraints (for example, that the white king is closer to
the white rook than the black king).

The predicate $\Seq$ denotes a sequence of strategy moves of white
followed by legal moves of black. Again,
$\underset{i\in\{0..k\}}{\Seq}\ \varI{p}_i\ \varI{m}_i\ \varI{p}_i'\
\varI{p}_{i+1}$ is just a shorthand notation:

$$
\underset{i\in\{0..k\}}{\Seq}\ \varI{p}_i\ \varI{m}_i\ \varI{p}_i'\
\varI{p}_{i+1} \equiv \bigwedge_{i \in \{0..k\}}
\left(\rel{\wsI}^{\varI{m}_i}\ \varI{p}_i\ \varI{p}_i' \ \wedge\ \M_i\
  \varI{m}_i \ \wedge\ \rel{\bI}\ \varI{p}_i'\ \varI{p}_{i+1}\right)
$$

Recall that the parameter $\varI{m}$ in the $\rel{\wsI}^\varI{m}$
relation denotes the type of the move played (e.g.,
$\rel{\wsI}^{\ImmediateMateMove}\ \varI{p}\ \varI{p}'$ denotes that
there is an immediate mate in position $\varI{p}$, leading to the
position $\varI{p'}$). Predicates
$\M_i$ additionally constrain the strategy move types played by white
(e.g., some $\M_i$ can require that $\varI{m}_i$ belongs or does not
belong to some set of move types).

Finally,
$\underset{i\in\{0..k\}}{\Post}\ \varI{p}_i\ \varI{m}_i\ \varI{p}'_i$
is just a shorthand for a postcondition that relates all positions and
move types encountered during such a sequence of moves, i.e.,
$$
\underset{i\in\{0..k\}}{\Post}\ \varI{p}_i\ \varI{m}_i\ \varI{p}'_i
\equiv \Post\ \varI{p}_0\ \varI{m}_0\ \varI{p}_0'\ \varI{p}_1\ \ldots
\ \varI{p}_k\ \varI{m}_k\ \varI{p}_k'\ \varI{p}_{k+1}
$$

In most of useful statements, the postcondition is just a relation between
the starting and the ending position i.e., it is of the form
$\Post\ \varI{p}_0\ \varI{p}_{k+1}$.

Some variations of the above general form are allowed, and the series
can end either by a black move (as given above), or by a white move.

Statements of the above form can express that some invariant is
preserved or that some measure decreases after a series of moves (so
it is a termination measure), or that some kinds of moves cannot or
must be played after a series of moves, or that some series of moves
leads to checkmate, and so on.

Note that we have assumed that lemmas are expressed in terms of the
strategy relation $\rel{\wsI}^{\varI{m}_i}$. However, we could use the
function $\dfun{\wsI}^{\varI{m}_i}$ if it is
available and we want to reason directly about it.

All central lemmas used in several available correctness proofs for
Bratko's KRK endgame strategy \cite{Bratko78,icga,krk-cade} fit into
the above form. For example, one of those lemmas claims that after a
\emph{basic move} ($\SqueezeMoveTxt$, $\ApproachMoveTxt$ or
$\KeepRoomMoveTxt$), only a basic or a \emph{mate move}
($\ReadyToMateMoveTxt$, $\ImmediateMateMoveTxt$) can be played.

\begin{eqnarray*}
  \forall \varI{p}_0\,\varI{p}_0'\,\varI{m}_0\,\varI{p}_1\,\varI{p}_1'\,\varI{m}_1.&& \neg \WRcaptured{\varI{p}_0}\ \wedge\\
                                           && \rel{\wsI}^{\varI{m}_0}\ \varI{p}_0\ \varI{p}_0' \wedge \varI{m}_0 \in \BasicMoves \wedge \bI\ \varI{p}_0'\ \varI{p}_1 \wedge \\
                                           && \rel{\wsI}^{\varI{m}_1}\ \varI{p}_1\ \varI{p}_1'\\ 
                                           && \longrightarrow \varI{m}_1 \in \BasicMoves \vee \varI{m}_1 \in \MateMoves
\end{eqnarray*}

\paragraph{Methods for proving lemmas}
Although, in principle, statements of the above form can be proved
manually (usually, only informally), we are interested in fully or
semi-automated proofs within some computer system. Therefore, the
statements have to be given in terms of the endgame definitions. Note
that $\Pre$, all $\M_i$, and $\Post$ should have an executable
implementation.

\begin{itemize}
\item Since the set of positions $\pI$ and the set of available move
  types are finite, then this is a finite-domain conjecture that can
  be checked by an exhaustive enumeration. This approach can be more
  efficient if the functions $\ndfun{\wsI}$ and $\ndfun{\bI}$ defined
  in Section \ref{sec:endgamestrategies} are available so not all
  possible positions $\varI{p}_i$ must be considered, but only their
  subsets. Additionally, if the strategy is given by a function
  $\dfun{\wsI}$, then the search space (the quantification domain) is
  smaller (branching occurs only with the black moves). This can
  naturally be implemented in a language that enables iterating over
  sets of positions (so, it is simple in C or in \IsabelleHOL, but not
  in URSA).

\item Such statements can also be considered as constraint
  solving problems and constraint solvers can be used to show that there are
  no positions and move types that would violate the implication,
  i.e., to show that
$$
\Pre\ \varI{p}_0\ \wedge\ \underset{i\in\{0..k\}}{\Seq}\ \varI{p}_i\
\varI{m}_i\ \varI{p}_i'\ \varI{p}_{i+1} \wedge \neg
\underset{i\in\{0..k\}}{\Post}\ \varI{p}_i\ \varI{m}_i\ \varI{p}'_i
$$
\noindent is unsatisfiable. It can be formulated in a language
suitable for a constraint solver, given the relations $\rel{\wsI}$ and
$\rel{\bI}$, and the conditions $\Pre$, $\M_i$ and $\Post$ (not
necessarily effectively executable) can be formulated in the language
of the constraint solver. In our case, the formula was either a
bitvector-arithmetic (BVA) formula that was bit-blasted to SAT (via
URSA) and solved by a SAT solver, or was a linear integer arithmetic
(LIA) formula formulated within \IsabelleHOL{} that was converted to
SMT representation (via \IsabelleHOL) and solved by an SMT solver.
\end{itemize}

\paragraph{Gluing lemmas together} Unlike lemmas, the central theorem 
often cannot be expressed in the language of a decidable theory. Namely, 
proving the central theorem may require, for instance, some inductive 
argument and using undecidable theories. In some computer-based proving 
approaches (e.g., in a general-purpose programming language or in some 
constraint systems) inductive arguments cannot be expressed, so the 
proof of the central statement must remain informal. On the other 
hand, proof assistants allow such forms of reasoning, so the central 
statement and its proof (using the lemmas) can be rigorously expressed 
and mechanically verified within the system. 

\subsubsection{A Case Study of the Strategy for KRK}
In order to prove the central theorem, i.e., that the strategy for
white is winning, we need to prove that it is terminating and
partially correct. In this section we list all lemmas that together
show that the strategy is correct (they are similar to lemmas given in
\cite{icga}, but somewhat improved, thanks to facts discovered while
formally proving that they can be glued together).

\begin{lemma}
  After a strategy move by white, black cannot capture the white rook.
\end{lemma}

\begin{lemma}
  After an $\ImmediateMateMoveTxt$ move, black is checkmated.
\end{lemma}

\begin{lemma}
\label{lemma:readytomate}
  $\ReadyToMateMoveTxt$ move leads to checkmate in the next move.
\end{lemma}

\begin{lemma}
  $\RookHomeMoveTxt$ and $\RookSafeMoveTxt$ are played only in the
  first three moves.
\label{lemma:rookhome-safe}
\end{lemma}

\begin{lemma}
  Starting from a position with a room greater than 3, playing three
  full moves where white plays only basic strategy moves
  ($\SqueezeMoveTxt$, $\ApproachMoveTxt$ or $\KeepRoomMoveTxt$) reduces the
  room or leaves the room the same, but decreases the Manhattan
  distance between the white king and the critical square.
\end{lemma}

\begin{lemma}
  When the room is less or equal to 3, after a three full moves where
  white plays only basic strategy moves, the next move must be a
  mating move ($\ReadyToMateMoveTxt$ or $\ImmediateMateMoveTxt$).
\end{lemma}

All six lemmas can easily be formally stated in the general form
described previously. For example, Lemma \ref{lemma:readytomate} can
be formalized as follows:

\begin{eqnarray*}
  \forall \varI{p}_0\varI{p}_0'\varI{m}_0\varI{p}_1\varI{p}_1'\varI{m}_1.&& \neg \WRcaptured\ \varI{p}_0\ \wedge\\
                               &&  \rel{\wsI}^{\varI{m}_0}\ \varI{p}_0\ \varI{p}_0' \wedge \varI{m}_0 = \ReadyToMateMove \wedge \bI\ \varI{p}_0'\  \varI{p}_1 \wedge \\
                               &&  \rel{\wsI}^{\varI{m}_1}\ \varI{p}_1\ \varI{p}_1' \\
                               &\Longrightarrow&  \varI{m}_1 = \ImmediateMateMove \wedge \checkmateI\ \varI{p}_1'
\end{eqnarray*} 

We formulated the same set of above lemmas in three different systems:
C, URSA and \IsabelleHOL. 

In \IsabelleHOL, encoding the lemmas is rather straightforward. For
example, Lemma \ref{lemma:readytomate} is formulated as follows:

\begin{isar_code}
{\tt
\begin{tabbing}
  \hspace{5mm}\=\hspace{5mm}\=\hspace{35mm}\=\kill
  theorem ReadyToMateMove: \\
  \>  "$\forall$ p0 p1 p1' p2 t2.\ \=$\neg$ $\WRcaptured$ p0 $\wedge$ \\
  \>\>      $\StrategyWhiteMove$ p0 p1 $\ReadyToMateMove$ $\wedge$  \\
  \>\>      $\legalmoveBK$ p1 p1' $\wedge$ \\
  \>\>      $\StrategyWhiteMove$ p1' p2 t2 $\longrightarrow$ \\
  \>\>\>        t2 = $\ImmediateMateMove$ $\wedge$ $\checkmateI$ p2"\\
  \end{tabbing}
}
\end{isar_code}

Lemmas are formulated over the record-based representation, and are
converted to the pure LIA and proved by applying SMT
solvers. Translation is done manually, but could be automated by
implementing a suitable tactic.

In URSA, the quantification is implicit and the statement is proved by
showing that the negated statement is unsatisfiable:

{\footnotesize
\begin{verbatim}
call LegalPositionWhiteToMove(nPos1w, bLegalWhite1);
call IsRookCaptured(nPos1w, bRookIsCaptured1);
call StrategyRelation(nPos1w, nPos1b, bRel1, nsReadyToMate);
call LegalMoveBlack(nPos1b, nPos2w, bBlack1);
call StrategyRelation(nPos2w, nPos2b, bRel2, nStep2);
call Mate(nPos2b, bMate);
assert(bLegalWhite1 && !bRookIsCaptured1 && bRel1 && bBlack1 && bRel2 && !bMate);
\end{verbatim}
}

In C, we used the brute-force enumeration-based approach to prove the
lemmas. It required using nested loops that correspond to
quantification (we won't further discuss this C-based approach).

Expressing the lemmas is simple in each of the above approaches. In
the enumeration-based approach in C, proving lemma is actually
execution of the code that expresses it. In \IsabelleHOL, the user
must provide a proof, but that proof just needs to provide a
boilerplate code that instructs the system to transform the lemma into
the SMT-LIB form, run the external SMT-solver (we used the Z3 solver),
import the answer and, possibly, verify it (in the case when
proof-reconstruction is required and when the answer contains the
proof-certificate). In URSA, proving lemma is automatically delegated
to the underlying SAT solver by transforming the high-level
specification into a propositional formula.  The specifications using
three approaches (C, URSA, and Isabelle) are rather short, even
including formulations of lemmas -- the C file has around 1000 lines,
the URSA file around 2000 lines, and the Isabelle files around
7800. Table \ref{tbl:function} (and Table \ref{tbl:summary} in a
summarized form) shows statistics (the number of variables and clauses
in the generated SAT instance and the running time) for verifying the
specification that uses the deterministic strategy function (with no
optimizations) using the constraint solver URSA.\footnote{The
  specification and the lemmas are slightly changed compared to the
  earlier version \cite{icga}, thanks to the influence of combination
  of different proving approaches. The size of formulae and the
  running times are also slightly different.}

\begin{table}[!ht]
\begin{center}
\noindent\begin{tabular}{@{}lrrrrrrr@{}} \hline
            & Lemma 1 & Lemma 2 & Lemma 3 & Lemma 4 & Lemma 5 & Lemma 6 & Total \\ \hline
variables   & 38778   & 38935   &  77552  &  146180 &  117039 &  146198 &  564682 \\
clauses     & 150911  & 151795  & 302173  &  575404 &  455909 &  575645 & 2211837 \\
time        &    56s &      6s &      20s &    6107s&   2060s &       36s &  8285s \\ \hline
\end{tabular}
\end{center}
\caption{Deterministic strategy function with no optimizations (the URSA approach)}
\label{tbl:function}
\end{table}

\section{Steps Beyond: Citius, Altius, Fortius}
\label{sec:advanced}

After proving correctness of the basic version of the strategy using
different approaches, we want to make steps beyond, trying to perform
proofs {\em faster}, to make them even {\em higher}-level, and to
reformulate conjectures to be more general and {\em stronger}.  We
will go through a series of iterations using high-level statements and
the constraint solver URSA. The culmination of that is a machine
verifiable correctness proof (in \IsabelleHOL) for the strategy
generalized to an $n \times n$ board, for arbitrary natural number
$n$.\footnote{The strategy and the proof can be generalized to
  $m \times n$ case, but we will stick to the $n \times n$ case since
  it is more interesting as it admits using more symmetries in
  reasoning.}

\subsection{Citius: Faster Computations and Efficiency Issues}
\label{subsec:citius}

There are several ways to make checking and proving {\em faster}.

\paragraph{Using underspecifications}
As discussed in Section \ref{sec:endgamestrategies}, a
non-deterministic strategy is underspecified, and its deterministic
version has to give specific choices for each strategy move (there
must be a way to choose if there are more than one move meeting the
conditions). If the retrograde analysis is used, then it is faster to
verify the deterministic strategy definition, than the
non-deterministic one. However, the deterministic version becomes more
difficult for analysis by high-level conjectures (as it is more
complex). In addition, if correctness of the non-deterministic
strategy has been previously shown, then it is sufficient to show that
the deterministic version just refines its non-deterministic
counterpart. This brings us to a more general and subtle idea, used
often in interactive theorem proving (and actually used in our
\IsabelleHOL{} and URSA proofs). We can reason about strategy moves,
not only using their concrete definitions, but using only their
preconditions and postconditions (that do not necessarily cover all
details of the moves), as announced in Section
\ref{sec:endgamestrategies}. This way, our conjectures may be more
general and much easier to verify (both in the constraint solving and
in the interactive proving setting). While the executable function is
capable of computing the single move that is played in each legal
position and can effectively be used in a chess playing system, the
strategy expressed in terms of a relation is much more general and
covers various possible refinements.

For example, in the strategy in a form of a function, we used the {\em
  maximal} $\SqueezeMoveTxt$ move -- one that maximally reduces the
room for the black king. This can enable white to more quickly reach
mate in some cases, but it  affects the proving process. In the
non-deterministic version, we use a loosely specified
$\SqueezeMoveTxt$ (guaranteed to reduce the room, but not necessarily
maximally).

The proving process is significantly more efficient when the relation
is used, but the determinism is lost.  If correctness of the
deterministic function is still to be considered, then the total time
should include time for the proof that the function meets the
requirements of the strategy relation -- 638 seconds. The total
proving time is, therefore, 1382s (744s+638s) which is significantly
smaller than 8285s, when only the function is considered (Table
\ref{tbl:summary}).

In the following, we will introduce some further optimizations,
leading to new versions of both the non-deterministic and the
deterministic strategy.

\paragraph{Using equivalent specifications}
One of our main goals in developing formalizations of chess endgames
(within different systems) is to have easily understandable,
high-level descriptions of relevant notions and strategy moves, and
all definitions used so far were designed with this goal in mind.
However, they are often expressed in a way that, when unfolded,
produces huge corresponding formulae. Consider, for instance, a
formalization of mate in URSA:

{\footnotesize
\begin{verbatim}
procedure Mate(nPos, bMate) {
  call LegalPositionBlackToMove(nPos, bLegalPositionBlackToMove);
  call BKCannotMove(nPos, bBKCannotMove);
  call WRAttacksBK(nPos, bBKAttacked);
  bMate = bLegalPositionBlackToMove && bBKCannotMove && bBKAttacked;
}
\end{verbatim}
}

The above description is elegant and easily understandable. However,
when unfolded, it produces a huge corresponding formula. It can be
described in a much more focused way. Namely, white can actually mate
black only within a few patterns (for example, in the KRK endgame, the
black king must be on one of the four edges). So, instead of the above
general and readable definition, we can describe these patterns, as
given in the following URSA specification (note that the condition
that the position is legal can be dropped, since all mating positions
are explicitly represented):

{\footnotesize
\begin{verbatim}
procedure MateOpt(nPos, bMate) {
  call Bitvector2Pos(nPos, nWKx, nWKy, nBKx, nBKy, nWRx, nWRy, bWhiteOnTurn);
  call AbsDiff(nBKy, nWRy, nBKynWRy); call AbsDiff(nWKy, nBKy, nWKynBKy);
  call AbsDiff(nBKx, nWRx, nBKxnWRx); call AbsDiff(nWKx, nBKx, nWKxnBKx);
  bMate = !bWhiteOnTurn &&
   (nBKx == 0 && nWRx == 0 && nWKx == 2 && nBKynWRy > 1 && 
    ((nBKy != 0 && nBKy != 7) || nWKynBKy <= 1) && 
    ((nBKy == 0 || nBKy == 7) || nWKy == nBKy)) || 
   (nBKy == 0 && nWRy == 0 && nWKy == 2 && nBKxnWRx > 1 && 
    ((nBKx != 0 && nBKx != 7) || nWKxnBKx <= 1) && 
    ((nBKx == 0 || nBKx == 7) || nWKx ==n BKx)) || 
   ...
}
\end{verbatim}
}

The above specification generates a smaller formula, easier to digest
by the solving process. However, there are still two major
issues. First, while the former specification
is simple, understandable and very likely without errors, the latter
is complex and prone to errors. Here, the idea of {\em refinement} helps 
again: one first makes a specification in the style of the former one, 
then a specification in the style of the latter one, and then
{\em checks} (using a constraint solver or a proof
assistant) that these two are {\em equivalent}. Once this is done, in
the further proving process, only the more efficient one can be
used. Such process of refinement is often used in interactive theorem
proving, but here we show that it is also applicable within constraint
solving systems (and possibly in other proving approaches). The second 
issue is how the (optimized) specification can be derived
in the first place. The answer is: by a careful analysis and again
using computer support. One can start by a first incarnation of the
optimized specification and check if it matches the high-level one. If
it does not, then a computer system (for instance, URSA) provides
instances where the two specifications do not match, the user fixes
those differences and iterates the process. This is often demanding,
but also rewarding at the end: one has simplicity, understandability,
reliability, and efficiency.

The described approach was applied also to the stalemate definition,
yielding additional speed-ups.

The refinement approach was used in simplifying the strategy in one
more way: strategy moves specify what conditions should hold after the
move, but not necessarily what piece white should move. For example,
it can be proven that black can be mated only by the white rook, and
not by the white king, and this fact can be used to simplify some
definitions.

The alternative characterizations of mate and stalemate within URSA,
gave a speed-up (Table \ref{tbl:summary}). Note that the total time
should also include the 0.3s needed to prove that two specifications
for mate and that two specifications for stalemate are equivalent. It
is not necessary to verify that the function meets the requirements of
this newer version of the strategy relation -- since the two
relation-based specifications are proved to be equivalent.

\smallskip

Conditions that a certain move cannot be played in a given position,
which are basic building blocks for the strategy relation definition
(described in Section \ref{sec:krkstrategy}), can also be further
refined and optimized (again carefully proving the equivalence with
the original formulations). For example, the $\noImmediateMate$
condition can be optimized by noticing that a mate move can be played
only by moving the rook to an edge (where the black king is). 

Table \ref{tbl:summary} shows the gain of using the optimized NoMove
conditions (except for the $\noSqueezeMove$ condition that will be
discussed in Section \ref{subsub:nxn}).  Again, the total proving time
should include proofs that original definitions are equivalent to the
optimized ones: under 5s altogether.

\paragraph{Exploiting symmetries}
Chessboard has a number of symmetries and this can be exploited to
$(i)$ make the definitions simpler and more readable and $(ii)$ to
make reasoning more efficient. Symmetries in chess endings have
already been studied, for example by Bain \cite{Bain94learninglogical}.

There are three basic symmetries -- horizontal, vertical, and
diagonal. Reflection functions map squares to squares: if $\filesA$
and $\ranksA$ are global constants (denoting the numbers of files and
ranks)\footnote{Later we will also consider boards of other sizes
  than $8 \times 8$, so we consider symmetries in a bit more general
  framework.} equal to 8, then the horizontal reflection
$\mathcal{R}_h$ maps a square $(x, y)$ to the square
$(\filesA-x-1, y)$, the vertical reflection $\mathcal{R}_v$ maps it to
$(x, \ranksA-y-1)$ and the diagonal reflection $\mathcal{R}_d$ to
$(y, x)$ (the diagonal reflection is applicable only if the board is
square). If a position is specified by the squares assigned to pieces
on the board, then its reflected image is obtained by applying the
reflection function to all those squares.

We define that a KRK position that has the black king on the square
$(\bkf, \bkr)$, the white king on the square $(\wkf, \wkr)$, and the
white rook on the square $(\wrf, \wrr)$ is in {\em canonical form} if
the triple $(2\cdot \bkf+1, 2\cdot \wkf+1, 2\cdot \wrf+1)$ is
lexicographically smaller or equal to the triple
$(\filesA, \filesA, \filesA)$, if the triple
$(2\cdot \bkr+1, 2\cdot \wkr+1, 2\cdot \wrr+1)$ is lexicographically
smaller or equal to the triple $(\ranksA, \ranksA, \ranksA)$, and the
triple $(\bkf, \wkf, \wrf)$ is lexicographically smaller or equal to
than the triple $(\bkr, \wkr, \wrr)$ (if the rook has been captured,
then the third components are ignored). Essentially, in canonical
positions the black king is confined to be in a triangle that covers
approximately one eighth of the board and the other two pieces are
relevant only on the boards with odd dimensions, when the black king
is on the central square or on the diagonal. If the black king is on
the diagonal, then the white king must be on the diagonal or below it,
and if it is also on the diagonal, then the rook must be on the
diagonal or below it.

It is easy to define a function that checks if a position is in
canonical form. Reflections can be used to map any position into 
canonical form and a canonization function (e.g., the procedure 
\texttt{Canonize} in URSA) can be easily defined as a composition
of reflections.  
Canonization can be used to simplify definitions. For example,
in URSA, the procedure \texttt{MateOpt} that we have previously
shown can be simplified to the following one.

{\footnotesize
\begin{verbatim}
procedure MateOptSym(nPos, bMate) {
  call Canonize(nPos, nPosC);
  call Bitvector2Pos(nPosC, nWKx, nWKy, nBKx, nBKy, nWRx, nWRy, bWRCaptured, bWhiteOnTurn);
  call AbsDiff(nBKy,nWRy,nBKynWRy); call AbsDiff(nWKy,nBKy,nWKynBKy);
  call AbsDiff(nBKx,nWRx,nBKxnWRx); call AbsDiff(nWKx,nBKx,nWKxnBKx);
  bMate = !bWhiteOnTurn &&
      (nBKx==0 && nWRx==0 && nWKx==2 && nBKynWRy > 1 &&
           (nBKy!=0 || nWKynBKy <= 1) && (nBKy==0 || nWKy==nBKy)) ||
      (nBKy==0 && nWRy==0 && nWKy==2 && nBKxnWRx > 1 && 
           (nBKx!=0 || nWKxnBKx <= 1) && (nBKx==0 || nWKx==nBKx));
}
\end{verbatim}
}

The stalemate definition in URSA is also very succinctly expressed
using symmetries and canonization. Some other predicates can also be
reformulated in such manner, yielding a more readable
formalization. Table \ref{tbl:summary} shows the result of
reformulating the definitions of mate, stalemate and
$\ReadyToMateMoveTxt$ step.  Note that this does not have a
significant effect on the proving efficiency, but the main gain lies
in readability and in conciseness of the specification. Again, the
total time should also include 0.4s needed to prove that
specifications are equivalent.

Symmetries and canonical positions can also significantly improve
efficiency if they are used for the so called \emph{without loss of
  generality (wlog) reasoning} \cite{wlog}. The central lemma for
exploiting symmetries (formally proved in \IsabelleHOL) states
that if there is some property of chess positions invariant under all
three kinds of reflections (if the property holds for a position, then
the same property holds for its reflected image), then, in order to
show that all positions satisfy that property, it suffices to show
only that canonical positions satisfy that property.

\begin{isar_code}
{\tt
\begin{tabbing}
\hspace{5mm}\=\hspace{5mm}\=\kill
{\bf theorem} symmetry:\\
\>  {\bf fixes} P :: "KRKPosition $\Rightarrow$ bool"\\
\>  {\bf assumes} \="$\forall$ p. P (reflectx\_p p) $\longrightarrow$ P p"  \\
\>\>          "$\forall$ p. P (reflecty\_p p) $\longrightarrow$ P p"   \\
\>\>          "$\forall$ p. P (reflectdiag\_p p) $\longrightarrow$ P p"\\
\>  {\bf assumes} "$\forall$ p. is\_canon p $\longrightarrow$ P p" \\
\>  {\bf shows} "$\forall$ p. P p"
\end{tabbing}
}
\end{isar_code}

We have proved in \IsabelleHOL{} that all relevant notions are
invariant under all three types of reflections. For example, if the
black king is checkmated in a given position, it is also checkmated in
its reflected image (e.g. $\checkmateI$ (reflectx\_p p)
$\longleftrightarrow$ $\checkmateI$ p).

Since many notions are used, it is a tedious job to formulate all such
lemmas in a proof-assistant, but once they are formulated, they are
all almost trivial to prove (and almost all proofs can be obtained
automatically). Therefore, any statement that has an
  outermost quantifier that universally quantifies over all positions
can be relaxed by adding the condition that the position is
canonical, which significantly reduces the search space.

Exploiting symmetries is formally justified in \IsabelleHOL{} (all
described lemmas have been formally proved), while in URSA and C
approaches it is used without justification within the system. Using
the wlog symmetries in the URSA approach, led to additional
significant speed-up shown in Table \ref{tbl:summary}.

Since this is the final version in our chain of refinements, in Table
\ref{tbl:relationOpt} we present its detailed statistics.

\begin{table}[!ht]
\begin{center}
\noindent\begin{tabular}{@{}lrrrrrrr@{}} \hline
            & Lemma 1 & Lemma 2 & Lemma 3 & Lemma 4 & Lemma 5 & Lemma 6 & Total \\ \hline
variables   & 28190   & 27964   &  55958  & 111946  & 85186   & 111947  & 421191 \\ 
clauses     & 98948   & 98166   & 196467  &  393049 & 299472  & 393060  & 1479162 \\ 
time        & 3s      & 3s      & 6s      &  218s   &  72s    & 14s     & 316s \\ \hline
\end{tabular}
\end{center}
\caption{Final version of the non-deterministic strategy}
\label{tbl:relationOpt}
\end{table}

With all the presented optimizations, we built a new (final) version
of the strategy function and again proved all
the lemmas. The running times for proving lemmas are presented in
Table \ref{tbl:functionOpt}, showing significant speed-up compared to
the initial version of the deterministic strategy.

\begin{table}[!ht]
\begin{center}
\noindent\begin{tabular}{@{}lrrrrrrr@{}} \hline
            & Lemma 1 & Lemma 2 & Lemma 3 & Lemma 4 & Lemma 5 & Lemma 6 & Total \\ \hline
variables   &  33574  &  33718  &  67097  &  129518 & 101359  & 129536  & 494802 \\ 
clauses     & 116391  & 116792  & 232535  &  450829 & 351951  & 450962  & 1719460 \\ 
time        & 14s     & 3s      & 7s      &  2244s  &  1566s  & 15s     & 3849s \\ \hline
\end{tabular}
\end{center}
\caption{Final version of deterministic strategy}
\label{tbl:functionOpt}
\end{table}

The presented optimizations affect the definition of the deterministic
strategy as the optimized predicates are used both in the relation and
the function definition. Since it was proved that the original
function refines the basic strategy relation and that all
reformulations were justified, the basic function also refines the
final, optimized relation. Also, it can be separately proved that the
optimized version of the function refines the optimized relation. This
lemma (with 44668 variables and 154471 clauses) is proved in 559s,
reducing the overall proving time for the function from 8285s, to
316s+559s=875s (plus 6s in total for proving the lemmas that justify
the optimizations). This is still significantly less than 3849s needed
for directly proving the lemmas for final deterministic version, which
illustrates the power of the refinement used.

A summary of the effect of different optimizations in the URSA
specification is given in Table \ref{tbl:summary}.

\begin{table}[!ht]
\begin{center}
\begin{tabular}{@{}lrrr@{}} \hline
                                  & variables  & clauses  & time\\ \hline
Function (direct)                 & 563872     & 2208153  & 8285s \\
Function (via relation)           & 626270     & 2448464  & 744+638s \\
Relation                          & 394096     & 1480708  & 744s \\
optimizing Mate and Stalemate     & 380577     & 1374585  & 668s + 0s \\
optimizing NoMove                 & 372987     & 1357965  & 569s + 5s \\
reformulations using symmetry     & 420987     & 1478400  & 524s + 0s \\
wlog reasoning                    & 421191     & 1479162  & 316s \\ 
Optimized function (direct)       & 494802     & 1719460  & 3849s \\
Optimized function (via relation) & 465859     & 1633633  & 316s + 559s \\ \hline
\end{tabular}
\end{center}
\caption{Summary of the proving process for different version of URSA specification}
\label{tbl:summary}
\end{table}

\paragraph{Automation and efficient solvers/theories}
Communication with external SAT/SMT solvers significantly increases
automation in \IsabelleHOL. However, this requires formulating
conjectures in appropriate theories. As we already noted, we were able
to formulate all central lemmas in the language of linear arithmetic
and this enabled their efficient, automated proofs using the Z3 solver
integrated with \IsabelleHOL. This required to reformulate the
definition of room to avoid multiplication, as discussed in Section
\ref{sec:ourbratko}. Otherwise, the theory of bit-vector arithmetic
would have to be used, leading to less efficient proofs.

\subsection{Altius: High-level Proofs and Understandability Issues}

The understandability of the correctness proofs comes from the
formulation of the central high-level lemmas. For example, in our
current proof, there is a lemma that claims that $\RookHomeMoveTxt$
and $\RookSafeMoveTxt$ moves can be played only in the first three
moves (Lemma \ref{lemma:rookhome-safe}). We have demonstrated that
this lemma can be formally expressed and proved fully automatically
(for example, by using SMT solvers). However, this lemma can be
replaced by three simpler ones. The first one claims that the
$\RookSafeMoveTxt$ can be played only as a first move (the
$\RookSafeMoveTxt$ cannot be played immediately after any strategy
move of white, followed by a legal move of the black king). The second
one claims that $\RookHomeMoveTxt$ can be played only immediately
after $\RookHomeMoveTxt$ or $\RookSafeMoveTxt$. The third one claims
that $\RookHomeMoveTxt$ cannot be played immediately after two
$\RookHomeMoveTxt$ moves. These three lemmas together imply our
original lemma, but also give us a higher-level understanding and more
insight into the strategy details. Additionally, it turns out that it
is much faster to prove three simpler lemmas: in URSA, the original
lemma is proved in around 218s in the fastest variant, while the three
simpler lemmas together require only around 58s. One reason for that
is that the original lemma is too coarse and requires reasoning about
7 plies at the same time, while the simpler lemmas require reasoning
about only up to 5 plies at the same time.

Further insights can be obtained by formulating explicit moves'
preconditions, postconditions and invariants. Namely, in all previous
approaches it remains only implicit what relationship between the
pieces has been established after the first three moves, and what has
exactly happened after the first, after the second, and after the
third move. Finding explicit characterizations (expressed only in
terms of positions and not the strategy) is a complicated task, but it
would make the proof more understandable and would significantly
contribute to deeper understanding of the very finest details of the
strategy.

For example, $\RookSafeMoveTxt$ can be played only when no other move
can be played. This condition is complicated (it takes into account
the definitions of all moves in our strategy). However, by inspection
and analysis of the positions in which this move can be played, we
managed to characterize those positions explicitly by the following
condition (shown in \IsabelleHOL).

\begin{isar_code}
{\tt
\begin{tabbing}
\hspace{5mm}\=\kill
      (l\=et \=$(\varI{\varI{WKx}}, \varI{WKy}) = \WK{\varI{p}}; (\varI{BKx}, \varI{BKy}) = \BK{\varI{p}}; (\varI{WRx}, \varI{WRy}) = \WR{\varI{p}}$;\\
\> $\varI{C_{BR}} = \chebyshev\ (\BK{\varI{p}})\ (\WR{\varI{p}}); \varI{C_{BW}} = \chebyshev\ (\BK{\varI{p}})\ (\WK{\varI{p}})$ \\
        \>in \= $($\=$\varI{C_{BR}} = 1 \wedge \varI{C_{BW}} = 2 \wedge \neg \WRDivides\ \varI{p}\ \wedge$\\
        \> \>\> $\varI{WRx} \neq \varI{WKx} \wedge \varI{WRy} \neq \varI{WKy} \wedge \varI{WKx} \neq \varI{BKx} \wedge \varI{WKy} \neq \varI{BKy})\ \vee$\\
        \> \> $(\varI{BKx} = 0 \wedge \varI{BKx} = 0 \wedge \varI{WKx} = 0 \wedge \varI{WKy} = 2 \wedge \varI{WRx} < 2 \wedge \varI{WRy} > 2)\ \vee$\\
        \> \> $(\room\ \varI{p} = 2 \wedge \varI{WKy} = 2 \wedge (\varI{WKx} = 0 \vee \varI{WKx} = 2)))$"
\end{tabbing}
}
\end{isar_code}

The first disjunct characterizes the positions where the white rook
must escape towards a far edge as all other moves would leave it
exposed, and the second and the third conditions characterize the
positions where black must escape towards edge not to leave the black
in a stalemate position (the third condition characterizes only two
very special positions where the black king is confined to only a
single square). A simple lemma is proved that claims that after any
strategy move followed by a legal move of black  
the above condition for $\RookSafeMoveTxt$ cannot be satisfied, so
$\RookSafeMoveTxt$ can be played only as a first move (when the
postconditions of all moves are examined, it is almost obvious why
this is so). This lemma ensures that $\RookSafeMoveTxt$ can be played
only in the first move, but this time we have a rather explicit
explanation and this proof could be done even manually.

Next, a lemma is proved that claims that after any two $\RookHomeMoveTxt$
moves the following condition holds:\footnote{This condition is recognized 
by Bratko \cite{Bratko78} (however, without the $\neg\, \incheckI\ \varI{p}$ 
condition that turns out to be necessary).}
$$\neg \WRExposed\ \varI{p}\,\wedge\, \WRDivides\ \varI{p} \,\wedge\, \neg \incheckI\ \varI{p} \,\wedge\, \room\ \varI{p} > 2$$

We follow Bratko's informal proof and prove that the following
condition is preserved by all moves of black and all moves of white
(except $\ReadyToMateMoveTxt$ and
$\ImmediateMateMoveTxt$):\footnote{In the white-to-move positions, the
  $\LPattern\ \varI{p}$ condition should actually be
  $\LPattern'\ \varI{p}$, where $\LPattern'$ is slightly changed
  condition $\LPattern'$ \cite{Bratko78}.}
$$\neg\, \WRExposed\ \varI{p} \,\wedge\, (\WRDivides\ \varI{p} \,\vee\, \LPattern\ \varI{p}) \,\wedge\, $$
$$\neg\, \incheckI\ \varI{p} \,\wedge\, \room\ \varI{p} > 2$$

The former condition ensures that a basic or a mating move can be
played. 

It turns out that lemmas that use such explicitly formulated pre and
post conditions and invariants are much easier to prove than lemmas
formulated only in terms of moves -- all these lemmas with explicit
invariants are together proved in under 10s in URSA which is a very
significant speed-up compared to the 218s needed for the original
lemma. Again, such explicit conditions bring not only speed but
higher-level understandability.

To conclude, pen-and-paper proofs (e.g., the one given by Bratko
\cite{Bratko78}) required simpler lemmas, as complex lemmas are hard
to prove manually.  On the other hand, formulating just a few very
coarse lemmas leads to a simpler proving process (although such lemmas
can require higher proving time on the computer, they require less
human time and effort to formulate them, which is the most significant
and most time consuming component), so there is a trade off between
the proof understandability and its efficiency.

\subsection{Fortius: Stronger Conjectures and Scalability Issues}

Having a conjecture and its proof at hand, we can consider if we can
prove a {\em stronger}, more general conjecture. For instance, we can
notice that the correctness of retrograde analysis is valid not only
for chess, but for a wide class of games that can be defined as a
loosely axiomatized theory. In this section we will focus on another
sort of generalization --- we will prove correctness of our KRK strategy
for generalized, $n \times n$ chessboards.

\subsubsection{Generalization to $n \times n$ Chessboards}
\label{subsub:nxn}

Although the standard chess game is played on a $8 \times 8$ board, we
can consider the KRK endgame and the presented strategy on
$n \times n$ boards. This generalization, made in the spirit of
mathematical generalizations, breaks the connection with the classic
chess game, but illustrates power of the presented proving
methodologies and also how they can be used for different games.

We easily modified our programs to be able to represent boards of
other sizes than $8 \times 8$. Namely, bitvectors used for
the board representation in URSA and in C can easily be adapted to a
variable board size, by changing the number of bits used to represent
coordinates. For example, in URSA, conversions from and to bitvectors
can be modified to include the dimensions of the board. 

\paragraph{Modifying the strategy}
Rapid testing of this generalized version using the C program quickly
revealed that the basic strategy does not define moves for all
positions in the $4 \times 4$ and $5 \times 5$ cases. We made two
adjustments to the strategy to cover those two cases.

First, in the $\ApproachMoveTxt$ and $\KeepRoomMoveTxt$ moves of the
original strategy it is required that the white king does not move to
an edge in order to keep the king out of the edge where the black king
is on, but, on small boards the white king might touch other
edges. This does not make any problems, so we reformulated this
condition and explicitly required that two kings are not on the same
edge.

Second, the original strategy forbids making a $\RookSafeMoveTxt$ move
such that the Chebyshev distance between the two kings is exactly two
(unless they are both next to the white rook). This is needed to keep
the black king from approaching the white rook in the next move, so
the white rook would need to run to a safe position again. However, in
several positions on the $4\times 4$ and $5\times 5$ boards it is not
possible to make such $\RookSafeMoveTxt$ move (so it is not possible
to make any move, since $\RookSafeMoveTxt$ is the last possible move
of our strategy). Because of that, a new strategy move
$\RookSafeSmallBoardsMoveTxt$ (non-existent in the original Bratko's
strategy) had to be introduced at the end of the current strategy and
it needs to be used (in certain positions) only in the $4 \times 4$
and $5 \times 5$ cases:

\begin{itemize}[leftmargin=\parindent]
\item [{\bf 8.}]  $\RookSafeSmallBoardsMoveTxt$: If none of the above
  is possible, then move the rook to an edge where the white king is
  (if not already on that edge); in the reached position, the
  Chebyshev distance between the white rook and the white king has to
  be 2.
\end{itemize}

The modified strategy is correct for all board sizes $n \geq 4$ (the
additional move kind will be played only when $n=4$ or $n=5$).

We  measured the total proving time needed for different
chessboard dimensions and the results are summarized in Table
\ref{tab:C-URSA-n}. The C program used a retrograde analysis, and URSA
was used to prove the lemmas about the correctness of the most
optimized version of the relation.\footnote{It is interesting that the
  total size of formulae for the lemma and the time needed was greater
  for $n=15$ than for $n=16$. This is because of the representation
  used: for $n=16$, four bits are used for coordinates and the
  conditions for ensuring that the coordinates are within the board
  disappear, contrary to, for instance, the case $n=15$.}  Again, if
correctness of the final strategy function is considered, then the
total proving times by URSA for each $n$ should include proofs that
the function refines the final strategy relation. This additional time
for $n=4$ is 8s, for $n=7$ exceeds the time used for proving lemmas --
316s, and for $n=11$ exceeds our time limit -- 1h.

\begin{table}[!ht]
\begin{center}
\begin{tabular}{@{}rrrrr@{}} \hline
$n$ & No. of legal positions & plies to win &   C    & URSA  \\ \hline 
4   &     1312               &  21          &    0s  &   37s \\ 
8   &   175168               &  65          &    5s  &  316s \\ 
12  &  2360160               & 109          &  194s  & 1250s \\ 
16  & 14241920               & 153          &  847s  & 2904s \\  \hline 
\end{tabular}
\end{center}

\caption{CPU time (in seconds) required by the C approach and the URSA 
approach for proving strategy correctness for dimensions from $n=4$ to 
$n=16$}
\label{tab:C-URSA-n} 
\end{table}

\paragraph{Reformulating the NoSqueeze condition}
While the retrograde analysis proved its usefulness in preliminary
experimenting (for example, in showing incompleteness of the
generalized strategy for small boards), Table \ref{tab:C-URSA-n} shows
that its running time grows quickly as $n$ grows. The URSA and
Isabelle approaches also becomes practically unusable for higher
dimensions. 

Meeting the limit in proving correctness for larger dimensions by any
of the approaches needs new deep insights. For instance, a strategy
move like $\SqueezeMoveTxt$ is played by the rook and, within the
strategy description, all possible 14 moves by the rook are
covered. For the $8 \times 8$ board, this approach does not do any
harm. However, for the $1000 \times 1000$ case, the specification
involves $999 + 999$ possible rook moves, this condition explodes and,
together with other similar conditions, makes the conjecture
impossible to resolve in a reasonable time.

If the maximal $\SqueezeMoveTxt$ cannot be played, then no
$\SqueezeMoveTxt$ at all can be played. A deeper analysis reveals that
any maximal $\SqueezeMoveTxt$ move fits into one of only 16 patterns,
independent of the dimensions of the chessboard. Assume that the
position of the white king, the white rook, and the black king after
the $\SqueezeMoveTxt$ move are respectively $(\wkf, \wkr)$,
$(\wrf, \wrr)$, $(\bkf, \bkr)$. The rook must not be exposed after the
move of white, so it must hold that
$\max(|\wkf - \wrf|, |\wkr - \wrr|) \leq \max(|\bkf - \wrf|, |\bkr -
\wrr|)$.
The maximal $\SqueezeMoveTxt$, if it can be achieved, is played only
in case of equality, i.e., if one of the following holds:
$2 \wrf = \wkf + \bkf$, $2 \wrr = \wkr + \bkr$,
$\wrf + \wrr = \bkf + \wkr$, $\wrf + \wrr = \wkf + \bkr$,
$\wrf - \wrr = \bkf - \wkr$, $\wrf - \wrr = \wkf - \bkr$. Expressing
$\wrf$ and $\wrr$ gives all candidate positions for the maximal
$\SqueezeMoveTxt$. If $2 \wrf = \wkf + \bkf$, then
$\wrf = (\wkf + \bkf) / 2$, but if $\wkf + \bkf$ is odd, then
$\wrf = (\wkf + \bkf + 1) / 2$ is played. Also, if
$\wrf + \wrr = \bkf + \wkr$, then either $\wrf = \bkf + \wkr - \wrr_0$
and $\wrr = \wrr_0$ or $\wrf = \wrf_0$ and
$\wrr = \bkf + \wkr - \wrf_0$, where $(\wrf_0, \wrr_0)$ is the
position of the white rook before $\SqueezeMoveTxt$. Also, in some
maximal $\SqueezeMoveTxt$ moves, the white rook moves to the file or
rank next to one of $\wrf = \bkf + 1$, $\wrf = \bkf - 1$,
$\wrr = \bkr + 1$, $\wrr = \bkr - 1$. This gives 16 candidate
positions.

We used the above approach in URSA and Isabelle/HOL and we
have shown equivalence between the optimized and the original
definitions.
  
On the $8\times 8$ board, the number of 16 potential positions for
$\SqueezeMoveTxt$ is larger than the number of all possible positions
for the rook (the rook can potentially move to one of the 14
squares). Therefore, we excluded this reformulation from our
experiments for the $8 \times 8$ chessboard. However, as the dimension
of the board rises, the number of possible moves by the rook
increases, but the number of maximal $\SqueezeMoveTxt$ candidate
positions remains the same. This is vitally important and enables us
to efficiently reason about arbitrarily large boards, as it
turns out that it is easy to characterize all move types with a number
of possible candidate positions that does not depend on the board size
(for $\ImmediateMateMoveTxt$ there are 4 candidate positions, for
$\ReadyToMateMoveTxt$ there are 12, for $\SqueezeMoveTxt$ there are
16, $\ApproachDiagMoveTxt$, $\ApproachNonDiagMoveTxt$,
$\KeepRoomDiagMoveTxt$, and $\KeepRoomNonDiagMoveTxt$ are played by
the king so there are 8 candidate positions for them, and for both
$\RookSafeMoveTxt$ and $\RookHomeMoveTxt$ there are only 4 candidate
positions for the rook).

This reformulation not only makes a significant leap in scalability
and efficiency of the proving, but it is also well-suited for
communicating the strategy to human players (as they must consider a
smaller number of candidate moves).

\paragraph{Making the board size arbitrary}
The above approach made proving strategy correctness for large
dimensions possible. But, still, these proofs are proofs for concrete,
individual dimensions and not for arbitrary dimension. So, now we need
another deep insight: all conditions used in specification of the
strategy are expressed in terms of linear arithmetic,  not only in
terms of coordinates of the pieces, but also if the dimension is
treated as a variable (and not as a constant)! This observation brings
us to a general correctness theorem expressible in terms of linear
arithmetic and provable by a proof assistant equipped with a support
for SMT solving (for linear arithmetic). On the other hand, this
technique is not applicable within the SAT-based URSA approach.

Therefore, when using the system equipped with support for reasoning
in linear arithmetic, we can have a single theorem that shows that the
strategy is correct for a chessboard of {\em any} size $n \times n$,
for $n \geq 4$. The times to prove the lemmas for this theorem for the
strategy relation in \IsabelleHOL{} are shown in Table
\ref{tbl:isabellenxn}.

We show the times for using Z3 in the oracle mode (without the proof
reconstruction) and also in the fully verified mode.\footnote{The
  reported times are for Isabelle2015. The new SMT proof
  reconstruction module introduced in Isabelle2016 is significantly
  less efficient in our case than the older one, since it is tailored
  towards simpler and smaller proofs that occur in typical sledghammer
  tasks.} Note that the proof reconstruction consumes most of the time
(but also gives a strongest guarantees). The time for verifying the
whole theory (everything except those six lemmas, including gluing
them together) is around 259 seconds.

\begin{table}[!ht]
\begin{center}
\noindent\begin{tabular}{@{}lccccccc@{}} \hline
         & Lemma 1 & Lemma 2 & Lemma 3 & Lemma 4 & Lemma 5 & Lemma 6 & Total \\ \hline
oracle   & 1s      & 1s      & 2s      & 76s     & 38s      & 18s     & 136s   \\ 
verified & 2s      & 2s      & 3s      & 803s    & 816s     & 576s    & 2202s  \\ \hline
\end{tabular}
\end{center}
\caption{Isabelle proofs for symbolic $n$ }
\label{tbl:isabellenxn}
\end{table}

\section{Related Work}
\label{sec:related}

We are not aware of another complex conjecture proved using three 
different computer-supported approaches. However, combining reasoning 
tools and approaches is present for decades and is used in a number of 
contexts and application areas. In the following text, we make just 
a brief selection of such works.

SAT solving has been used for solving very hard mathematical
combinatorial problems like the Boolean Pythagorean triples problem
\cite{HeuleKM17}. There are integrations of FOL provers with SAT
solvers \cite{BiereDKV14}.
SAT solvers can be used for solving CSP problems 
\cite{TamuraTB10,CadoliMP06,StojadinovicM14} and also
SAT/SMT solvers can be combined with constraint programming 
solvers \cite{Stuckey10,Bankovic16}.
Computer algebra systems have been plugged into SAT solver to provide 
a system that can be used as an assistant in proving process for 
either finding counterexamples or finitely verifying universal 
conjectures. This system has been used for proving a number 
of complex mathematical conjectures such as conjectures from graph 
theory regarding properties of hypercubes \cite{ZulkoskiBHKCG17}.

SAT solvers were used from the Isabelle system in proving the
Erd\"os-Szekeres conjecture for convex polygons with at most 6 points
\cite{Maric-Erdos}. SMT solver are interfaced to interactive theorem
provers \cite{BlanchetteBP13, ArmandFGKTW11, PengG15}
and successfully used for complex tasks, such as verification 
of analog-mixed signal circuits \cite{PengG15-verification}.
FOL provers were used from the Isabelle system in formalizing 
Tarski's geometry \cite{DurdevicNJ15}.

A combination of constraint programming and theorem proving was used
for software verification \cite{CollavizzaRH10}.  Combination of
interactive and automated proving for FOL was used for reasoning about
lazy functional programs \cite{BoveDS12}.  Constraint programming is
applied at the test suite reduction problem \cite{GotliebCLMP16}.
There are systems that combine testing and interactive theorem prover
to reason about programs and to automatically generate concrete
counterexamples \cite{abs-1105-4394}.

Experimental mathematics is an approach to mathematics in which
computation and experimentation are used to investigate mathematical
objects and suggest conjectures, properties and patterns.
Experimental mathematics is frequently based on general purpose
computer algegra systems and on custom built software. There are
research journals focused on this approach to mathematics, such as the
journal {\em Experimental Mathematics}.

There is also a long history of computer based analysis of games and
strategies. Computers have been often used for constructing chess
endgame databases.  Early programs for retrograde analysis were
implemented by Thompson \cite{thompson-lookup} and Bramer
\cite{Bramer80,Bramer1982}.  Databases for a number of chess endgames
are publicly available (e.g., the Lomonosov Endgame Tablebases,
generated by Zakharov and Makhnichev, contain optimal play for all
endgames with seven or less pieces).

Recently, computers have also been used to reason about endgame
database correctness (e.g., to construct endgame databases that are
correct by construction). Reasoning is based mainly on retrograde
analysis and enumerations of positions and moves. Machine verifiable
proofs of database correctness for chess KRK endgame database were
given by Hurd \cite{hurd2005,HurdH09}. A combination of binary
decision diagrams (BDDs) for representing positions, model checking
tools for automation, and the proof-assistant HOL for high assurance
were used.

Other games were also analyzed using computers. For example, Schaeffer
et al.~showed that checkers game (on an $8 \times 8$ board) is a draw
\cite{Schaeffer2007}. Their argument included a giant endgame table,
obtained by using massive computations combined with sophisticated
search algorithms. Neither a high-level nor a machine verifiable proof
was produced. Edelkamp applied BDDs to two-player games to improve
memory consumption for reachability analysis and game-theoretical
classification \cite{Edelkamp02symbolicexploration}. Gasser
\cite{Gasser96solvingnine} showed that the game of Nine Men's Morris
is a draw, using a combination of endgame databases and search.

Only a very few high-level endgame strategies are accompanied by
correctness proofs. For instance, Zuidema \cite{Zuidema1974} and
Morales \cite{Morales,Morales96} didn't prove correctness of their
strategies for KRK. On the other hand, Bratko gave an informal,
pen-and-paper proof for his KRK endgame strategy \cite{Bratko78}.
There are only few direct, computer-supported correctness proofs of
strategies for chess endgames. A SAT-based constraint solver URSA was
used by Malikovi\'c and Jani\v{c}i\'c to prove correctness of a KRK
strategy closely related to the one considered in this paper
\cite{icga}. Machine verifiable proofs were not provided and the
lemmas cannot be glued together into a single theorem. A similar
approach was used by Mari\' c, Jani\v ci\' c and Malikovi\' c in
\IsabelleHOL{} and automated SMT solvers were used to automatically
prove the conjectures \cite{krk-cade}, leading to a self-contained,
fully machine verifiable proof of the strategy correctness. We are not
aware of other specifications of chess strategies within a proof
assistant or a constraint programming system (the aforementioned work
on verifying chess endgame databases \cite{hurd2005,HurdH09} does not
deal with strategies understandable to humans).

\section{Conclusions}
\label{sec:conclusions}

In this paper we have shown how one can use computer support for
proving correctness of a chess endgame strategy, and we advocate that
the same methodology could be used for proving some other complex
combinatorial conjectures (over finite domains). We have presented a
case study of one particular problem and shown how one can reason
about chess in a rigorous mathematical manner, supported by
state-of-the-art computer tools. We have shown that KRK chess endgame
is {\em strongly solved} \cite{Schaeffer2007}, i.e. it is win for
white for $n \times n$ chessboard, for each natural numbers $n$
greater than $3$. We revisit some key questions and list some of the
main lessons that we learnt and that can be used for proving other
combinatorial conjectures.

\paragraph{Specification language and verifiability} 

\begin{itemize}
\item \emph{Proofs developed within proof assistants have the highest
    degree of confidence.} Representing the general chess rules and
  the endgame in \IsabelleHOL{} leads to highest possible reliability.

\item \emph{Proof assistants have better expressive power than
    alternatives.}  For instance, we could not glue the lemmas
  together (we could not even express the central theorem) in C and
  URSA, but we did do it in \IsabelleHOL.

\item \emph{Formalization should include an executable implementation
    of the analysed algorithm.} Defining an executable strategy
  function and relation enabled making various experiments (e.g., URSA
  made possible to find all positions that satisfy some criteria), but
  also directly enabled some reasoning methods (e.g, the retrograde
  analysis). Verified executable strategy formalization can be
  exported to a general purpose programming language.
\end{itemize}

\paragraph{Convincibility, faithfulness, and understandability}

\begin{itemize}
\item \emph{Use a small set of basic definitions, separate basic and
    auxiliary definitions.} The basic definitions are cornerstones in
  any problem formalization and they must be concise, clear, and
  carefully manually inspected since the proofs are checked only
  modulo these. In our study, we invested a lot of effort in building
  a simple, understandable problem representation (in different
  settings) based only on general chess rules.

\item \emph{Introduce derived notions to speed up reasoning, but
    formally show connections with basic definitions.}  We introduced
  many definitions specific for the KRK case, but we have always
  formally shown that they are in accordance with the general chess
  rules.

\item \emph{Introduce notions as abstractly as possible, so that they
    can be reused in other scenarios.} Identifying chess notions
  relevant for other two-player games can help analysing those games.

\item {\em Use a problem representation understandable to humans,
    whenever possible}. Instead of a endgame tablebase, we focused on
  a strategy represented in a form of an algorithm, usable both by
  humans and computers.

\item {\em Use symmetries as they can lead to more concise and
    understandable definitions, but also to more efficient
    reasoning}. Using symmetries should be justified by a meta-wlog
  theorem. We used symmetries for deriving optimized definitions of
  mate and other notions.

\item \emph{Clearly identify preconditions, postconditions and
    invariants.} The best understanding of algorithm comes by
  identifying the program state in all points of its execution, that
  can be described by lemmas that formulate preconditions,
  postconditions and invariants.

\item \emph{Optimized exhaustive search has both good and bad sides.}
  Retrograde analysis was a very efficient way to prove the strategy
  correctness both in C and Isabelle/HOL, but it did not provide us
  with better understanding, and proved to be inapplicable to large
  board sizes.
\end{itemize}

\paragraph{Abstraction, refinement, and generalization}
\begin{itemize}
\item \emph{Use non-deterministic specification whenever possible and
    introduce deterministic specifications only when necessary.} We
  first introduced strategy in form of a relation, and defined the
  strategy function only in the end.
  
\item \emph{Reason about the most abstract definition that guarantees
    correctness and only afterwards introduce specific implementation
    details.} It was much faster to prove correctness of the strategy
  specification that leaves questions of choosing the squeeze move
  open.

\item \emph{Try to make definitions independent of the problem
    dimensions.} Finding only a finite number of candidate positions
  for each move enabled us to prove the strategy correct for arbitrary
  large boards.
\end{itemize}

\paragraph{Automation and efficiency}
\begin{itemize}
\item {\em Whenever possible, express relevant notions and conjectures
    in terms of theories supported by efficient automated theorem
    provers (for SAT, SMT, FOL, etc).} We adapted all notions so that
  they fit into QF-LIA (e.g., we had to change Bratko's notion of
  room, and replace quantifiers by finite conjunctions), which enabled
  us to heavily use \emph{automation}: SMT solvers via Isabelle/HOL
  and SAT solvers via URSA. We relied on available automated
  procedures as much as possible, pushed them to their current limits
  and used human effort only for tasks requiring real insights and
  ideas.
 
\item {\em Proofs by automated theorem provers have to be verified by
    proof assistants.} Connection between the URSA endgame definition
  and the exported SAT representation is not formally shown (and
  relies on the correctness of the URSA system implementation). On the
  other hand, in \IsabelleHOL, transformation from the record-based
  representation to LIA can be implemented separately and LIA
  representation can then be automatically transformed into SMT-LIB
  (due to the SMT support in \IsabelleHOL).

\item {\em If a candidate lemma cannot be proved by some computer
    tool, it is beneficial to get its
    counterexamples}. Counterexamples can give crucial hints on how to
  fix errors, and such corrections took majority of our time and
  effort. We used SMT solver within \IsabelleHOL{} that can give one
  counterexample (typically -- a position that does not meet the
  statement), and -- URSA that can list them all.

\item {\em Use fast, even not maximally reliable tools for rapid
    testing of conjectures.}  Checking of lemmas can be time-consuming
  and can take minutes. In contrast, the retrograde analysis can check
  the overall correctness of the strategy in only seconds, and
  therefore is much more suitable for rapid testing of variants of the
  strategy. However, the retrograde analysis does not provide the
  explanations of why the strategy works when it does and why the
  strategy does not work when it doesn't.

\item \emph{There is a trade-off between the coarseness of lemmas and
    time to prove them.} Coarse lemmas were easy to formulate, but
  required long time to prove. Splitting them to smaller lemmas
  required much more effort but brings pay off in much faster automated
  proofs.

\item \emph{Save time in proof evolution, not in the final formal
    proving.} The final, central theorem, within the Isabelle system,
  was proved in a few minutes. However, this time is not very
  important: once the theorem is polished, the proof is generated and
  verified only once. What is critical is that the road to this
  theorem requires checking and proving many conjectures and it is
  critical that these steps can be performed reasonably fast, possibly
  only in terms of seconds or minutes, so the proving process is
  really interactive. We believe that the presented synergy of
  different approaches enable this {\em seeking for the proof}
  approach practically usable in mathematical practice.
\end{itemize}

Our proofs are an illustration for a successful synergy among
different computer-supported proving approaches. Our experience is
that computer tools available nowadays provide a strong and reliable
support for proving non-trivial conjectures from mathematics and
computer science. This modern support for formal reasoning is a big
step forward, analogous to a support for ground calculations that
computers made available decades ago. Most important attributes of
this modern support for reasoning are reliability and
automation. Still, this support is far from being able to replace
mathematician: it cannot provide intuition, deep insights, or proof
ideas and there is no magic computer button for proving complex
theorems. Rather, we show that one can use a mixture of proving
approaches, methods, tools, ideas, and tricks as extremely valuable
help in filling technical gaps in proofs and linking together a number
of arguments.

\section*{Acknowledgement}
The first and the second author are partly supported by the grant
174021 of the Ministry of Science of Serbia.  The authors are grateful
to the anonymous reviewers for their very detailed and helpful
comments on this paper.



\bibliographystyle{alpha}

\end{document}